\pdfoutput=1 
\documentclass[sigconf,9pt]{acmart}

\copyrightyear{2025}
\acmYear{2025}
\setcopyright{rightsretained}
\acmConference[SIGCOMM '25]{ACM SIGCOMM 2025 Conference}{September 8--11, 2025}{Coimbra, Portugal}
\acmBooktitle{ACM SIGCOMM 2025 Conference (SIGCOMM '25), September 8--11, 2025, Coimbra, Portugal}\acmDOI{10.1145/3718958.3750495}
\acmISBN{979-8-4007-1524-2/2025/09}



\usepackage{amsmath}
\usepackage{xcolor}
\usepackage{xspace}
\usepackage{listings}
\usepackage[detect-all]{siunitx}
\usepackage[capitalise]{cleveref}
\usepackage{algorithm}
\usepackage[noend]{algpseudocode}
\usepackage{bytefield}
\usepackage{tikz}
\usepackage{placeins}
\usepackage{pgfplots}\pgfplotsset{compat=1.18}
\usepackage{colortbl}
\usepackage{fontawesome}
\usepackage{pifont}
\usepackage{fp}
\usepackage{array}

\usepackage{enumitem}
\setlist[description]{wide=0pt,leftmargin=\parindent, topsep=2pt}
\setlist[itemize]{wide=0pt, leftmargin=\parindent}
\setlist[enumerate]{wide=0pt, leftmargin=\parindent}

\usetikzlibrary{shapes,arrows,positioning,calc,fit}
\usepgfplotslibrary{groupplots,statistics}

\newcommand{\fl}[1]{\ensuremath{\mathit{#1}}}
\newcommand{\resid}{\fl{ResID}\xspace}
\newcommand{\resinfo}{\fl{ResInfo}\xspace}
\newcommand{\bw}{\fl{BW}\xspace}
\newcommand{\In}{\fl{In}\xspace}
\newcommand{\Eg}{\fl{Eg}\xspace}
\newcommand{\strt}{\fl{StrT}\xspace}
\newcommand{\duration}{\fl{Dur}\xspace}
\newcommand{\sv}{\fl{SV}\xspace}
\newcommand{\dst}{\fl{DstAddr}\xspace}
\newcommand{\ts}{\fl{TS}\xspace}
\newcommand{\tg}{\mathrm{tag}\xspace}
\newcommand{\pkt}{\fl{pkt}\xspace}
\newcommand{\concat}{\;\|\;}

\newcommand{\keyop}[2]{#1\mathopen{}\left(#2\right)}
\DeclareMathOperator{\prfop}{\mathsf{PRF}}
\newcommand{\prf}[2]{\keyop{\prfop_{#1}}{#2}}

\DeclareMathOperator{\len}{len}

\DeclareRobustCommand{\field}[1]{%
\begin{tikzpicture}[baseline=(X.base)]
  \node[anchor=base,draw,inner xsep=1.5pt,inner ysep=0pt,minimum height=12pt] (X) {#1};
\end{tikzpicture}%
}
\renewcommand{\algorithmicrequire}{\textbf{Input:}}

\renewcommand{\algorithmicreturn}{Return:}
\newcommand{\droppkt}[0]{\texttt{drop\_pkt}}
\newcommand{\fwdbest}[0]{\texttt{fwd\_best\_effort}}
\newcommand{\fwdfly}[0]{\texttt{fwd\_flyover}}
\newcommand{\fwd}[0]{\textbf{fwd}}

\newcommand{\crefover}[1]{\ding{#1}~in~\cref{fig:intuition}}

\newcommand{\para}[1]{\smallskip\noindent\textbf{#1.}\thickspace}

\usetikzlibrary{arrows.meta}

\DeclareRobustCommand{\laptop}[0]{%
\begin{tikzpicture}[baseline]
  \node (N) [minimum size = 12pt, outer sep=0pt, inner sep=0pt,rectangle, fill=white] {};
  \node  [black, minimum size = 0pt, outer sep=0pt, inner sep=0pt,font={\LARGE}] at (N) {\faLaptop};
\end{tikzpicture}%
}

\tikzset{
    aslink/.style={color=black!50, ultra thick},
    aslinkhigh/.style={color=redfte, ultra thick},
    annotation/.style={redfte, pin edge={redfte}},
}

\DeclareRobustCommand{\server}[0]{%
\begin{tikzpicture}[baseline]
  \node (N) [minimum size = 9pt, outer sep=0pt, inner sep=0pt,rectangle] {};
  \node  [black!80, minimum size = 0pt, outer sep=0pt, inner sep=0pt] at (N) {\faServer};
\end{tikzpicture}%
}

\usepackage{graphicx}
\usepackage{xcolor}

\definecolor{greyfte}{HTML}{8b8b8b}

\definecolor{redfte}{HTML}{e41a1c}
\definecolor{bluefte}{HTML}{377eb8}
\definecolor{greenfte}{HTML}{4daf4a}
\definecolor{purplefte}{HTML}{984ea3}
\definecolor{yellowfte}{HTML}{ff7f00}

\colorlet{ctotal}{bluefte}
\colorlet{crequest}{redfte}
\colorlet{cresponse}{yellowfte}

\pgfplotsset{compat = 1.18}

\pgfplotsset{
    legend style = {font=\normalfont},
    tick label style = {font=\normalfont}
}

\pgfplotsset{dotted/.style={dash pattern=on 0pt off 2\pgflinewidth}}
\pgfplotsset{dashed/.style={dash pattern=on 2\pgflinewidth off 2\pgflinewidth}}
\pgfplotsset{dashdotdotted/.style={
    dash pattern=on 0pt off 2\pgflinewidth on 0pt off 2\pgflinewidth on 2\pgflinewidth off 2\pgflinewidth
    }
}
\pgfplotsset{dashdotted/.style={
    dash pattern=on 0pt off 2\pgflinewidth on 2\pgflinewidth off 2\pgflinewidth
    }
}

\tikzset{
}

\pgfplotsset{
    grid = none,
    every axis plot/.style={
        thick,
        line cap = round,
        mark options = {
            style = solid,
            fill = white,
            scale=0.8,
        },
    },
    every axis/.style={
        clip=false,
        cycle list name = bestnodash,
        legend style={draw=none},
        y label style={rotate=-90, at={(ticklabel cs:1.03)}, anchor=south west, font=\normalfont},
        x label style={at={(ticklabel cs:1.03)}, anchor=north east, font=\normalfont},
        height=5cm, width=\axisdefaultwidth,
        axis y line = left,
        axis x line = bottom,
        tick align = inside,
        axis line style = {-},  
        axis line shift=10pt,   
    },
    every tick/.style={
        major tick length = 0.1cm,
        minor tick length = 0.07cm,
    },
    every tick label/.style={
        font = \normalfont\rmfamily,
    },
}

\pgfplotscreateplotcyclelist{bestlist}{
    {bluefte, solid, mark = *},
    {redfte, dashed, mark = x},
    {yellowfte, dotted, mark = diamond*},
    {greenfte, dashdotted, mark = +},
    {greyfte, dashdotdotted, mark = square*},
    {purplefte, dashdotdotted, mark = square*},
}

\pgfplotscreateplotcyclelist{bestnodash}{
    {bluefte, solid, mark = *},
    {redfte, solid, mark = x},
    {yellowfte, solid, mark = diamond*},
    {greenfte, solid, mark = +},
    {greyfte, solid, mark = square*},
    {purplefte, solid, mark = asterisk},
}

\pgfplotscreateplotcyclelist{bestnomark}{
    {bluefte, solid, mark = none},
    {redfte, solid, mark = none},
    {yellowfte, solid, mark = none},
    {greenfte, solid, mark = none},
    {greyfte, solid, mark = none},
    {purplefte, solid, mark = none},
}

\pgfplotscreateplotcyclelist{shortlist}{
    {bluefte, solid, mark = o},
    {redfte, solid, mark = o},
}

\newcommand{\name}{\textsf{Hummingbird}\xspace}

\ifdefined\SuppressAnnotations
    \newcommand{\annot}[3]{}
    \newcommand{\todo}[1]{}
\else
    \newcommand{\annot}[3]{{\color{#1}{$\rule{8pt}{8pt}_{\textsf{\scshape\bfseries #2}}$ #3}}}
    \newcommand{\todo}[1]{\annot{red}{TODO}{#1}}
\fi

\lstdefinestyle{smallverb}{
  basicstyle=\ttfamily\footnotesize,
  columns=fullflexible,
  keepspaces=true,
}

\begin{document}

\title{\name: Fast, Flexible,  and Fair\texorpdfstring{\\}{ } Inter-Domain Bandwidth Reservations}
\fancyhead[LO]{Hummingbird}


\settopmatter{authorsperrow=4}
\newcommand{\symmysten}{\footnotemark[2]}
\newcommand{\symeth}{\footnotemark[3]}
\author{Karl Wüst}
\affiliation{%
	\institution{Mysten Labs}
	\country{}
}
\authornote{Both authors contributed equally to this paper}

\author{Giacomo Giuliari}
\affiliation{%
	\institution{Mysten Labs}
	\country{}
}
\authornotemark[1]
\author{Markus Legner}
\affiliation{%
	\institution{Mysten Labs}
	\country{}
}
\author{Jean-Pierre Smith}
\affiliation{%
	\institution{Mysten Labs}
	\country{}
}
\author{Marc Wyss}
\affiliation{%
	\institution{ETH Zurich}
	\country{}
}
\author{Jules Bachmann}
\affiliation{%
	\institution{ETH Zurich}
	\country{}
}
\author{Juan A. Garcia-Pardo}
\affiliation{%
	\institution{ETH Zurich}
	\country{}
}
\author{Adrian Perrig}
\affiliation{%
	\institution{Mysten Labs}
	\country{}
}

\renewcommand{\shortauthors}{Wüst et al.}

\begin{abstract}
	To realize the long-standing vision of providing quality-of-service (QoS) guarantees on a public Internet,
	this paper introduces \name: a lightweight QoS-system that provides fine-grained inter-domain reservations for end hosts.

	\name enables flexible and composable reservations with end-to-end guarantees, and addresses an often overlooked, but crucial, aspect of bandwidth-reservation systems: incentivization of network providers.
	\name represents bandwidth reservations as tradable assets, allowing markets to emerge. These markets then ensure fair and efficient resource allocation and encourage deployment by remunerating providers.
	This incentivization is facilitated by decoupling reservations from network identities, which enables novel control-plane mechanisms and allows the design of a control plane based on smart contracts.

	\name also provides an efficient reservation data plane, which
	streamlines the processing on routers and thus simplifies the implementation,
	deployment, and traffic policing, while maintaining robust security
	properties. Our prototype implementation demonstrates the efficiency and scalability of \name's asset-based control plane, and our high-speed software implementation can fill a 160 Gbps link with \name packets on commodity hardware.
\end{abstract}

\maketitle

\section{Introduction}

Most applications today require stable network connections to provide optimal functionality.
This holds true, in particular, for applications that rely heavily on the cloud or that require real-time communication.
Video calls, which have become more frequent with the popularity of remote work, are a common example of such applications.
Ensuring a high quality of service (QoS) for important calls can have a significant impact on the success of businesses.
In online games, the reliability of the connection between players and the game server can influence the outcome of a competition, providing an advantage to players with better network connectivity. The quality of the network connection can also impact traditional finance applications
as well as decentralized finance protocols~\cite{werner2021sok}. In both cases, ensuring that a trade is submitted on time can make or break the trade.
In B2B communication settings, QoS on a public Internet can satisfy the needs of high-availability use cases and replace the use of SD-WAN (valued at USD 5B in 2023~\cite{databridge-sdwan}) or leased line technologies (valued at USD 12B in 2023~\cite{verifiedmarketreports-leased-line}). Given the market sizes, incentives exist for challengers to leverage QoS to enter and capture market share.


To fully realize these use-cases, flexible, short-term guarantees for reliable network connectivity across multiple domains in the Internet are required.
However, in the current Internet, providing guaranteed QoS is only possible within a centralized domain, i.e., an autonomous system (AS), through expensive leased lines, or SD-WANs.
In addition, these guarantees cannot be flexibly obtained for short or medium time periods, i.e., on the order of seconds, minutes, hours, or days.

Previous proposals to provide QoS for Internet traffic~\cite{giuliari2021colibri,wyss2022protecting,giuliari2021gma,rfc2205,rfc2210,rfc2474} have several limitations. Earlier techniques~\cite{rfc2205,rfc2210,rfc2474} do not consider the presence of adversaries, and are therefore only applicable in intra-domain settings.
To withstand powerful network adversaries, more recent proposals~\cite{giuliari2021colibri,wyss2022protecting} enforce reservations through \emph{network capabilities}~\cite{anderson2004preventing}, unforgeable cryptographic tokens that can be checked efficiently and statelessly by routers on the forwarding path. However, the attack resilience of these systems comes at the expense of increased control-plane complexity~\cite{giuliari2021colibri}, or inflexible reservation sizes~\cite{wyss2022protecting,giuliari2021gma} (e.g., the source cannot specify the amount of bandwidth to be reserved).
The tradeoff between control-plane complexity and reservation flexibility may appear fundamental: Increasing the degrees of freedom in which reservations can be parameterized and composed \emph{should} come at the cost of a more intricate control-plane logic; conversely, a more streamlined control plane \emph{should} constrain reservation expressiveness.

This paper moves beyond this impasse by proposing \name, a system that allows expressive bandwidth reservations on the data plane, while avoiding the coordination and timeliness requirements that encumbered the control plane of previous proposals.

The key insights are that reservations can be made independent from in-band identities, such as network addresses, that they can be granted to the source on a per-hop basis without coordination between on-path ASes, and that they can be negotiated \emph{well ahead of their start time} if desired.
The source is then responsible for obtaining and composing reservations to protect end-to-end paths.

This decoupling of the control plane from the data plane enables new control-plane designs that address a fundamental part of a bandwidth-reservation system: A mechanism to incentivize fair and efficient allocation of reservations.

In contrast to previous solutions, we prioritize incentivization by designing a control plane that represents bandwidth as freely tradable assets---that can be sold, bought, and re-sold. This approach enables the creation of bandwidth markets that attach a monetary value to reservations, and ensure that bandwidth supply provided by ASes, and demand originating from hosts, converge to the fair market price.
This mechanism serves multiple purposes: (i) incentivize ASes to provide bandwidth reservations to hosts, (ii) maximize the utility of reservations and provide economic fairness, and (iii) ensure that the cost of preventing a host from obtaining a reservation is equal to the market value of the reservation.

We thus design a control plane that makes use of smart contracts to (i)~create tradable assets for bandwidth reservations, and (ii)~allow obtaining end-to-end QoS guarantees atomically without interaction between on-path ASes.

On the data plane, the sender adds authentication tags to each packet, serving as capabilities that are checked efficiently by border routers on the path.
\name incorporates several optimizations to ensure that reservations granted ahead of time cannot be abused, and that information about the heterogeneous on-path reservations can be transmitted efficiently.



\section{Related Work}\label{sec:relatedwork}

Several different systems to improve on best-effort traffic were proposed in the 1990s, of which the two most prominent are IntServ~\cite{rfc2205,rfc2210} and DiffServ~\cite{rfc2474}.
Unfortunately, neither of them is applicable to adversarial settings 
and both systems suffer from additional shortcomings:
IntServ can provide strong guarantees (in the absence of malicious actors) but requires a large amount of per-connection state and has to make complicated decisions on routers when processing RSVP requests~\cite{rfc2205}.
DiffServ, on the other hand, is highly scalable but only provides weak communication guarantees, and it is mainly used to define different traffic types \emph{within} an AS.
Thereafter, researchers have studied traffic prioritization protocols that use cryptographic authentication to enforce traffic prioritization~\cite{Yaar2004SIFF,anderson2004preventing,Yang2005,Parno2007,Lee2010Floc,Lee2013CoDef}. These systems, however, were either not scalable or did not provide sufficient guarantees compared to best effort, and never saw widespread adoption.

\para{Inter-domain bandwidth reservations}
The research area has seen a revival in the past decade with the rise of path-aware networking architectures, like SCION~\cite{chuat2022complete}, which facilitate bandwidth reservations with features such as packet-carried forwarding state and path stability~\cite{Hsiao2013,basescu2016sibra,giuliari2021gma,wyss2021secure,giuliari2021colibri,wyss2022protecting}.
However, even in this setting, previous proposals make different choices for the trade-off between scalability, deployability, fairness, simplicity, and security.
In the following, we describe two state-of-the-art systems and their limitations for our use case.

Colibri~\cite{giuliari2021colibri} provides its inter-domain bandwidth reservations with a two-step process. First, ASes establish segment reservations that cover SCION up-, core-, or down-segments.
Second, end-to-end reservations are established by combining such segment reservations, using fractions of the segment reservation bandwidth.
Establishing a Colibri reservation is performed by the source AS on behalf of the host using a \emph{gateway}, and requires coordination between all ASes on the path.
The gateway manages and monitors reservations for its hosts, which do not have access to the cryptographic keys to authenticate packets themselves. Instead, the gateway embeds the required authenticators in the packets.
Due to the complexity of separately monitoring each path, on-path ASes only monitor traffic probabilistically, and, if overuse is detected, punish the source AS, e.g., by declining future reservations.

Further, Colibri requires the deployment of an additional key distribution infrastructure (DRKey~\cite{kim2014lightweight,rothenberger2020piskes}), its control plane is complex to implement, and it requires duplicate suppression, i.e. filtering of duplicate packets, on the data plane. We discuss the need for duplicate suppression in related work and why \name can do without in \cref{sec:nodupsupp} in more detail.
Additionally, Colibri does not allow any partial reservations, in which bandwidth is only reserved on some of the hops, and thus can only be used if each AS on the path provides a reservation to the host.
Finally, Colibri offers little flexibility regarding the validity period of reservations:
a reservation is valid as soon as it is received and must be renewed after a fixed time interval of \SI{16}{\second}.

With Helia~\cite{wyss2022protecting}, Wyss et al.\@ introduce \emph{flyover} reservations, which are granted per-AS instead of the full path.
This reduces the complexity of reservations compared to Colibri and allows for partial reservations on a path -- an approach we adopt in \name as well.
One limitation of Helia is that the reservation granularity is coarse and, most importantly, the reserved bandwidth as well as the start and expiration times cannot be negotiated between the source and the on-path AS.
Instead, the reservations have fixed time slots and their size is automatically set to ensure that each source can obtain a reservation.
In addition, Helia does not allow end-hosts to create and use reservations directly. Similar to Colibri, it requires an AS-gateway to authenticate packets, it does not allow creating reservations ahead of time, and it also requires the DRKey~\cite{kim2014lightweight,rothenberger2020piskes} infrastructure to be in place.
Further, Helia does not support atomic end-to-end reservation guarantees, since the reservation for each hop is obtained individually without coordination.

\para{Network utilization and economic incentives}
Since the early days of the Internet, researchers explored the use of economic incentives to improve the utilization of network resources~\cite{cocchi1993pricing,mason1995pricingtheinternet}. These works were inspired by the recent memory of the ``congestion collapses''~\cite{jacobson1988congestion} of some years prior, and asked the question of whether pricing congestion \emph{explicitly} could help improve the allocation of the scarce bandwidth at the disposal of the early Internet. Kelly et al.~\cite{Kelly1998rate} demonstrate that TCP-like congestion control algorithms can naturally be interpreted as a price-based rate-allocation mechanism, where the ``shadow prices'' of are inferred from flow rates.
These publications served as the theoretical foundation for the later development of a rich literature on network utility maximization~(NUM) through implicit or explicit pricing~\cite{neely2013delay,akhil2019network,Fu2021learning}.

NUM-based systems have several key differences from \name.
First, they often consider a network with a single controller, or ``one-hop'' networks, where the controller can enforce the pricing and billing of resources. \name, instead, considers the full scope of inter-domain resource billing and enforcement, without a single central authority. Second, given the single-path nature of the Internet, these works only leave two options to the end-host: either to pay the price set by the network, or to reduce their network usage. \name builds on SCION's multi-path forwarding architecture, allowing end-hosts to reroute their traffic depending on congestion and prices.
Further, only reservation-protected traffic---which is a new and opt-in service--is priced through a market in \name{}. Therefore, our design does not alter the existing inter-domain economic relationships between ASes, and only adds a new layer of economic incentives for a new service.
\footnote{The emergent interactions between the incentive structures of the best-effort Internet and the reservation market are an interesting topic for future work.}

Finally, these works are mainly focused on incentive mechanisms, rather than concrete system design; \name, on the other hand, provides a complete design and implementation of a system that can be deployed in adversarial settings.
While our work considers a network and economic model that is very different from those studied in the NUM literature, analyzing \name under the lens of NUM is a promising direction for future work.

Regarding the use of blockchains as supporting infrastructure for network protocols, Route Bazaar~\cite{castro2015route} pioneered the use of blockchain to implement decentralized QoS markets.
Yet, this work does not provide a concrete system design, and does not consider reservation enforcement nor adversarial action.
\name, on the other hand, provides a complete design and implementation, including reservation monitoring and policing, preventing unauthorized access and overuse.

\section{\name Overview}\label{sec:overview}

In this section, we describe the properties and goals of \name, state the assumptions that we make on the network and infrastructure, and finally provide a high-level overview of the system design.

\subsection{Goals \& Properties}\label{sec:overview:properties}

The goal of \name is to enable end-to-end QoS guarantees in the Internet through bandwidth reservations.
Traffic using reservations is prioritized over best-effort traffic, and is therefore shielded from congestion and DoS attacks.

To that end, we rely on existing intra-domain traffic separation and prioritization mechanisms, e.g., DiffServ, or MPLS tunnels within an AS's network.
We focus on coordinating the utilization of these systems in the global inter-domain context, enabling their flexible and scalable composition, and ensuring that QoS-traffic does not suffer from congestion at peering points between ASes.

The core challenges we aim to solve are (i) to scale control and data plane to the size of the Internet; (ii) to avoid reservation spoofing and abuse by adversaries; and, (iii) to provide incentive mechanisms for the fair and efficient allocation of reservations.
In particular, we design \name to provide the following properties:

\para{Independent \& Composable Flyover Reservations}
Each reservation is granted for an individual AS hop, called a \emph{flyover}, and prioritizes the source's traffic (up to the reserved bandwidth) intra-AS from the ingress to the egress router of the AS, and inter-AS from the egress router to the ingress router of the next AS.
These reserved hops can then be composed to obtain end-to-end guarantees, or they can be used independently to only reserve parts of a path that are expected to be congested.
This composability also allows reusing a reservation on one hop for connections to different entities.
For example, one large reservation at a central AS in the network can be used together with multiple smaller reservations to obtain bandwidth guarantees for connections to multiple destinations.

\para{Control-Plane Independence}
The data plane is independent of the control plane; i.e., reservations can be created without relying on the network identity of the source.
This decoupling allows each AS to decide how it wants to offer reservations to its customers, and thus---in contrast to existing solutions---enables out-of-band mechanisms such as bandwidth marketplaces.

\para{Atomic Path Reservations}
Since reservations are obtained for each hop independently, it may occur that
a source can only obtain reservations for part of the path, while on a few key hops there is no bandwidth available.
If the source requires full path protection, then the obtained reservations are useless to the source, potentially cause a financial loss, and are no longer available to other hosts, despite being unused.
One useful property we require for \name's control plane is therefore a mechanism to provide \emph{atomic path reservations}, whereby the source is guaranteed to either obtain reservations for all requested hops, or none at all.

\para{Stateless Reservation Authentication}
Transit ASes can verify the authenticity of each reservation on the fly, based only on the reservation information in the packet. 

\para{Efficient Policing}
Monitoring and policing of bandwidth reservations in \name are feasible with simple methods and \emph{without} relying on duplicate suppression, which is challenging to implement in the network~\cite{lee2017case}.
Our design does not require complex probabilistic monitoring schemes, nor does it rely on punishment to prevent overuse.
Instead, traffic can be policed efficiently using deterministic monitoring and requires only storing minimal state.
Importantly, each AS can individually decide and limit the number of reservations that it can afford to monitor simultaneously.

\para{Reservation Incentives}
An important aspect of a bandwidth reservation is providing incentives for a fair and efficient resource allocation.
Since incentivization is usually an afterthought in the design of bandwidth-reservation systems, retrofitting such a mechanism onto the previous proposals creates significant inefficiencies and overhead.
\name is designed to enable efficient incentivization---through markets---with a control plane that represents bandwidth as a tradable asset (see \cref{sec:details:control}).

\subsection{Network \& Infrastructure}
\label{sec:overview:assumptions}

Our system requires the following minimal assumptions on network stability and infrastructure deployment.

\para{Path Stability \& Transparency}
Any bandwidth-re\-ser\-vation system requires the source to determine which ASes are traversed to obtain and use the correct reservations, and the paths need to remain stable through the duration of the reservation.
Therefore, we assume that the source is aware of the paths to the destination and their validity periods to be able to set up appropriate reservations.
This assumption is fulfilled, e.g., by path-aware network architectures that carry the path of packets in their headers such as SCION. As SCION is deployed commercially, with over a dozen ISPs selling connectivity\footnote{see \url{https://www.scion.org/isps/}}, we build \name on SCION.

\para{PKI for Autonomous Systems}
To ensure the validity of reservations, the control plane assumes that a PKI for ASes is in place, such as the Resource Public Key Infrastructure (RPKI)~\cite{rfc6480} used in secure BGP, or the Control-Plane PKI (CP-PKI)~\cite{chuat2022complete} used in SCION.

\para{Allocation of AS-internet resources}
We assume that ASes can allocate internal resources to support all reservations that they provide, i.e., traffic can pass between any two interfaces without causing congestion, as long as the advertised available bandwidth per interface is not exceeded.

\para{Time Synchronization}
To check the age of a packet, the clocks of hosts using reservations and ASes must be synchronized to a maximum clock skew of $\delta$ (e.g., \num{0.5} seconds). Several time synchronization systems amply satisfy this property, such as NTP, PTP, GNSS -- in practice, time accuracy with NTP is around 10ms, and below 100ms in the vast majority of cases~\cite{NTP-wikipedia}. A time synchronization error above \num{0.5}s can invalidate the QoS reservation.

\para{AS Stack}
ASes that provide reservations need to deploy a \name Service, which is responsible for coordinating the reservations and managing the AS-local secret keys for data plane authentication. 
Further, border routers need to be augmented to support the key derivation and packet authentication processes (see~\cref{sec:details:dataplane}).

\para{Client Stack}
\name clients must be able to (i) contact the control plane to obtain the reservation authentication keys, and (ii) use the keys to authenticate packets at forwarding time. 

\subsection{High-Level Design}

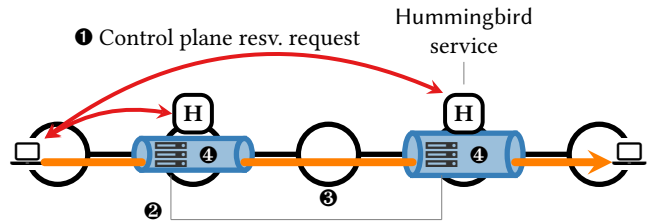
\begin{figure}[t]
	\centering
	\centering
\begin{tikzpicture}[
    asnode/.style={draw=black, fill=white, circle, minimum size=0.8cm, line width=2pt, inner sep=0pt, outer sep=0pt},
    aslink/.style={draw, line width=2pt, black},
    cparrow/.style={-stealth, redfte, line width=2pt},
    doublecparrow/.style={stealth-stealth, redfte, line width=2pt},
    numbernode/.style={black, minimum size = 0pt, outer sep=0pt, inner sep=0pt,font={\large}},
    csbox/.style={line width=1.5pt,anchor=center,draw,minimum size=0.5cm,inner sep=0pt, rounded corners=0.15cm, fill=white, font={\bfseries}},
    segr/.style={fill=bluefte!50, draw=bluefte, line width=2pt},
    segrlabel/.style={pos=0.5, anchor=north, black,opacity=1,font={\bfseries}},
    ]

    \node [asnode] (ASs) at (-5,0) {};
    \node [asnode] (ASx) [right = 1cm of ASs] {};
    \node [asnode] (ASw) [right = 1cm of ASx] {};
    \node [asnode] (ASz) [right = 1cm of ASw] {};
    \node [asnode] (ASd) [right = 1cm of ASz] {};

    \draw [aslink] (ASs) -- (ASx);
    \draw [aslink] (ASx) -- (ASw);
    \draw [aslink] (ASw) -- (ASz);
    \draw [aslink] (ASz) -- (ASd);

    \node (HS) [numbernode, fill=none] at (ASs.west) {\laptop{}};
    \node (HD) [numbernode, fill=none] at (ASd.east) {\laptop{}};

    \draw [segr] ([xshift=-8pt, yshift=6pt] ASx.west) rectangle ([xshift=8pt, yshift=-6pt] ASx.east);
    \draw [segr] ([xshift=-8pt] ASx.west) ellipse (3pt and 6pt);
    \draw [segr] ([xshift=8pt] ASx.east) ellipse (3pt and 6pt);
    \draw [segr] ([xshift=-8pt, yshift=8pt] ASz.west) rectangle ([xshift=8pt, yshift=-8pt] ASz.east);
    \draw [segr] ([xshift=-8pt] ASz.west) ellipse (3pt and 8pt);
    \draw [segr] ([xshift=8pt] ASz.east) ellipse (3pt and 8pt);

    \draw [line width = 3pt, yellowfte, line cap=round] ([yshift=-3pt]HS.east) to ([yshift=-3pt,xshift=-8pt]ASx.west);
    \draw [line width = 3pt, yellowfte, line cap=round] ([yshift=-3pt,xshift=8pt]ASx.east) to ([yshift=-3pt,xshift=-8pt]ASz.west);
    \draw [line width = 3pt, yellowfte, line cap=round, -stealth] ([yshift=-3pt,xshift=8pt]ASz.east) to ([yshift=-3pt]HD.west);

    \node [csbox] (CSx) at ([yshift=4pt]ASx.north) {H};
    \node [csbox] (CSz) at ([yshift=4pt]ASz.north) {H};
    \node [color=bluefte] (BRx) at ([xshift=3pt]ASx.west) {\LARGE\server{}};
    \node [color=bluefte] (BRz) at ([xshift=3pt]ASz.west) {\LARGE\server{}};
    \node (HSL) [yshift=0.8cm, anchor=south, align=left] at (ASz.north) {\name\\services};
    \draw [very thin, gray] (HSL) -- (CSx);
    \draw [very thin, gray] (HSL) -- (CSz);
    
    \draw [doublecparrow, bend left, line width=1.5pt] (HS.north east) to node [black, pos=0.45, anchor=west, yshift=0.2cm, xshift=-2.8cm, align=left] {\ding{202} Control plane\\resv. request} (CSz.north west);
    \draw [doublecparrow, bend left=20, line width=1.5pt] (HS.north east) to (CSx.west);

    \node (L2pos) at ([yshift=-20pt] BRx.south) {};
    \node (L2) [anchor=south east] at (L2pos) {\ding{183}};
    \node (L3) at ([yshift=-5pt] ASw.south) {\ding{184}};
    \node (L4a) [circle, fill=white, inner sep=-1.3pt] at ([xshift=5pt] BRx.east) {\ding{185}};
    \node (L4b) [circle, fill=white, inner sep=-1.3pt] at ([xshift=5pt] BRz.east) {\ding{185}};

    \draw [very thin, gray] (L2pos.north) -- (BRx.south);
    \draw [very thin, gray] (L2pos.north) -- (L2pos.north -| BRz.south);
    \draw [very thin, gray] (BRz.south) -- (L2pos.north -| BRz.south);


\end{tikzpicture}

	\caption{Overview of \name's operation on a path of five ASes (dark circles). The source host performs a reservation request (\ding{182}) to set up reservations on the path to the destination. Reservations may have independent sizes, start and expiration times (represented by the ``tube'' widths, \ding{183}). Further, not all on-path ASes \emph{must} provide reservations (\ding{184}). Border routers (\ding{185}) authenticate and prioritize reservation traffic on the data plane.}
	\label{fig:intuition}
\end{figure}

In the following, we provide an overview on how \name works, illustrated using  \cref{fig:intuition}, and how it provides the properties described in \cref{sec:overview:properties}.

At a high level, a flyover reservation is a six-tuple indicating which AS granted the reservation, the ingress and egress interfaces traffic will traverse, the start and expiration times, and the amount of forwarding bandwidth reserved.

\para{Obtaining Reservations}
To keep \name's operation simple and lower the burden on transit ASes, the source is responsible for obtaining valid overlapping reservations that cover the communication path on the control plane.

The control plane coordinates the creation of reservations, and allows establishing a shared reservation authentication key between the source and the transit AS.
Since reservations are granted on a per-hop basis, the control plane only requires pairwise interaction between the source and each individual on-path AS (\crefover{182}), and avoids complex multi-party reservation protocols.

In \name, reservations have a \emph{start time}. Thus, the reservation key can be provided to the source ahead of time, ensuring that the reservation is immediately usable, without any setup delays after the start time.

Since reservations are independent for each hop, reservations covering multiple hops do not need to have the same start time, expiration time, or even bandwidth (\crefover{183}).
Further, thanks to this decoupling, the source can obtain partial reservations on the path (\crefover{184}), even if not all on-path ASes provide reservations.

Our control plane provides incentivization and efficient utilization of resources by enabling \emph{bandwidth markets}, in which ASes can sell their reservation bandwidth by listing \emph{bandwidth assets}---which can be split, combined, and traded---that act as vouchers for reservations on the data plane. Once a source wants to use a reservation, it \emph{redeems} the asset to obtain the corresponding reservation authentication key.

In the bandwidth-reservation setting, where decentralization and resilience to failures are of paramount importance, a centralized bandwidth marketplace is unsuitable because it introduces a single point of failure.
Because of this, we design a control plane based on blockchain smart contracts that is decentralized, provides strong availability guarantees, and has a high performance.
This approach provides multiple advantages and is well-suited to the inherently decentralized setting of the Internet where not all parties mutually trust each other~\cite{wuest2018blockchain}. Blockchains are designed to be Byzantine fault tolerant (BFT) and thus highly resilient to failures and adversarial actions. They are replicated among globally distributed entities and thus provide redundant access to the \name control plane.

The use of smart contracts allows for flexible listing, modification, and purchase of assets with guaranteed integrity and without the interaction of the selling AS (see~\cref{sec:details:control}).
Further, smart contracts are executed atomically, i.e., state changes are only applied if the whole contract call executes without error, which enables sources to purchase reservations for a whole path in a single transaction, directly solving the problem of providing atomic path reservations.

It is important to note here that modern high-performance block\-chains such as Sui~\cite{blackshear2023sui}, which we use for our evaluation, do not suffer from often cited disadvantages that have plagued previous blockchain generations. They are far faster, cheaper, and much more energy efficient. 

\para{Using Reservations}
At forwarding time, the host uses the reservation keys obtained on the control plane to derive a per-packet message-authentication code (MAC) for each reserved AS hop.
Border routers in transit ASes can re-derive this reservation key on the fly based on the reservation information in the packet (\crefover{185}).
These cryptographic tags authenticate the reservation information and the packet's length, so that the reserved bandwidth can only be used by the source (see security analysis in \cref{sec:security}).

Further, reservations are not tied to flows. A source can use the same reservation for multiple flows crossing the same AS hop -- while ensuring that the flows \emph{in the aggregate} stay within the reserved bandwidth. \name's design allows policing reservation traffic to be lightweight, requiring only a single 8 byte counter per reservation (see \cref{sec:policing}).

\section{System Details}\label{sec:details}

This section presents the details of \name.
We first explain how reservations in our system are authenticated, then provide an example control plane that makes use of smart contracts and allows trading bandwidth assets in a market, and finally discuss the details of the data plane.

A full specification of the packet headers and format, based on the SCION Internet architecture~\cite{chuat2022complete}, is included in
\cref{sec:headerspec}.
In our specification, traffic that is (partially) protected by reservations uses a separate SCION path type that introduces additional 8~bytes per reserved hop compared to the standard SCION header.
We chose SCION as a basis for \name as it provides path transparency (an end host knows and can observe and select paths) and path stability (the packet's embedded path is followed), which are essential properties for bandwidth-reservation systems (see \cref{sec:overview:assumptions}). Moreover, SCION is deployed on commercial networks, enabling a roll-out of \name in practice.

While SCION offers native support for path awareness and cryptographically-enforced stability, the core of \name's design can be adapted to other architectures as well. For instance, IPv6 Segment Routing (SRv6)~\cite{SRv6} or MPLS-based~\cite{mpls} networks can support path transparency, and thus would also allow \name reservations.
However, these architectures also have limitations, compared to SCION, that make them less suitable for \name. SRv6 leverages IPv6 addresses, which are routed using BGP, and thus it is susceptible to the same transient rerouting events and hijacking attacks that prevent path stability in the current Internet.
MPLS-based networks, on the other hand, provide stable paths (unless the MPLS labels are announced and discovered through, e.g., BGP-LU~\cite{rfc3107}), but the configuration of MPLS network-to-network interfaces (NNIs) is still a complex and mostly manual process, limiting the scalability and flexibility of the system.
Thus, SCION remains the most practical target at present for realizing end-to-end enforceable QoS with market-based reservations.

\subsection{Reservation Authentication}\label{sec:details:authentication}

Assume that the source has reserved $\bw$ bandwidth from the ingress interface $\In$ to the egress interface $\Eg$ of an AS $K$, from time $\strt$ for a duration $\duration$.
AS $K$ assigns an identifier, $\resid$, to the reservation, which must be unique for the interface pair during the reservation's validity period.

The reservation is then determined by the tuple
\begin{equation}
	\resinfo_K = (\In, \Eg, \resid, \bw, \strt, \duration).\label{eq:resinfo}
\end{equation}
The AS $K$ providing the reservation is implicitly specified by the authentication key for the reservation, which is computed by AS $K$ as
\begin{equation}
	\label{eq:keyderivation}
	A_K = \prf{\sv_K}{\resinfo_K},
\end{equation}
where $\sv_K$ is a secret value known only to AS $K$, which is shared among its border routers.
$\mathsf{PRF}$ is a secure pseudo-random function with an output length sufficient to yield secure symmetric cryptographic keys (i.e., in practice at least \num{128} bits).
The authentication key, $A_K$, is shared through the control plane (see  \cref{sec:details:control}) with the source.

Note that, in contrast to previous bandwidth-reservation solutions, no source address or unique source identifier is used to create the reservation in our proposal.
Instead, a unique $\resid$ identifies the reservation at each hop.
This approach has several advantages:

First, each AS $K$ independently determines the $\resid$, which only has to be unique for a particular interface pair, and thus controls the maximum number of $\resid$s available.
This can improve the efficiency of monitoring and policing, which can be done deterministically (see \cref{sec:policing}).

Second, reservations can be shared between multiple (mutually trusting) sources or it can be obtained by one party for use by another.
For example, a client could obtain a reservation for the reverse direction (i.e., server to client) and provide the authentication key to the server in order to obtain bi-directional bandwidth guarantees (see \cref{sec:bidirectional}).

Further, the same source may have multiple res\-er\-va\-tions on the same hop.
This can increase the availability guarantees of the system (importantly, it can prevent attacks from on-reservation-set adversaries, see \cref{sec:adversarymodel}).
It also increases the flexibility of the system, e.g., during bursts of traffic, a source can obtain additional reservations to supplement an existing long-standing reservation.

Finally, it enables control-plane-independent reservations without introducing attacks that may be present if the reservations are granted per-source based on network identities.
If network identities would be used instead, the source would need to prove its identity when obtaining the reservation (e.g., through a PKI). Otherwise, an adversary could reserve bandwidth on the source's behalf to prevent that source from creating further reservations.

\subsection{Control Plane}\label{sec:details:control}


The \name data plane is compatible with any control plane that allows the source to (i) negotiate reservations and (ii) obtain $\resinfo_K$ and $A_K$ from AS $K$, importantly without relying on in-band identities. This control-plane independence enables the creation of a market for bandwidth reservations, which maximizes the utility of the allocated resources instead of relying on an in-band control plane.

Our control plane enables ASes to represent bandwidth as assets on a blockchain, makes them freely tradable, and allows obtaining end-to-end guarantees atomically. The core of our control plane is an \emph{asset contract} that defines the structure and behavior of \emph{bandwidth assets}.
These bandwidth assets are issued by ASes that provide \name reservations and act as vouchers for said reservations on the data plane.
Each asset represents reserved bandwidth for an ingress or egress interface at the issuing AS during a given time interval.
A pair of an ingress and an egress asset can then be redeemed at a later time for the information required to use the reservation on the data plane.

The asset contract is complemented by one or multiple independent \emph{market contracts} which provide a decentralized marketplace for trading bandwidth assets.
Assets can be split in the time or bandwidth dimension into multiple non-overlapping assets.
This allows an AS to issue a single ``large'' asset that represents all available bandwidth for a given interface for a long time interval, which can then be split by their current owner---which can be the AS, a buyer, or the market contract--- into ``smaller'' assets as needed.

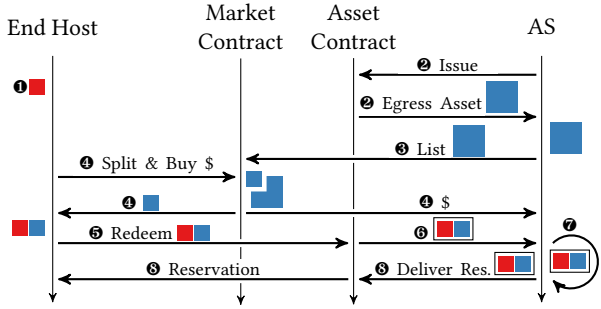
\begin{figure}[t]
	\centering
	{
	\centering
	\tikzstyle{rect}=[rectangle,draw=none,thick,fill=white,align=center,minimum height=0.75cm,minimum width={{width("Client")+14pt}}]
	\tikzstyle{smallasset}=[rectangle,draw=none,minimum size=6pt,inner sep=0]
	\tikzstyle{largeasset}=[rectangle,draw=none,minimum size=12pt,inner sep=0]
		\begin{tikzpicture}[every node/.style={transform shape},apply style/.code={\tikzset{#1}},>=stealth']
		\def\dist{0.8cm}
		\def\hdist{2.6cm}
		\def\slope{0}
		\def\spacing{2pt}
		\def\largeassetsize{12pt}
		\def\length{4.1}
		\node[rect] (endhost) {End Host};
		\node[rect, right=\hdist of endhost.center,anchor=center] (market) {Market\\Contract};
		\node[right=0.6*\hdist of market.center,rect,anchor=center] (assetcontract) {Asset\\Contract};
		\node[right=\hdist of assetcontract.center,rect,anchor=center] (as) {AS};
		
		\draw[->] (endhost.south) -- ($(endhost.south)-(0,\length*\dist)$);
		\draw[->] (assetcontract.south) -- ($(assetcontract.south)-(0,\length*\dist)$);
		\draw[->] (as.south) -- ($(as.south)-(0,\length*\dist)$);
		\draw[->] (market.south) -- ($(market.south)-(0,\length*\dist)$);
		
		
		\newcounter{nodeidx}\setcounter{nodeidx}{0}
		\foreach \i/\s/\l/\r/\t/\o/\v in {
			0.8/{<-,pos=0.5}/assetcontract/as/{\ding{183} Issue}/-1/1,
			1.5/{->,pos=0.35}/assetcontract/as/{\ding{183} Egress Asset}/1/1.3,
			2.2/{<-,pos=0.6}/market/as/{\ding{184} List}/-1.5/0.6,
			2.5/{->,pos=0.5}/endhost/market/{\ding{185} Split \& Buy \$}/1/1.5,
			3.1/{->,pos=0.65}/market/as/{\ding{185} \$}/-1/0.4,
			3.1/{<-,pos=0.4}/endhost/market/{\ding{185}}/-1/0.2,
			3.6/{->,pos=0.25}/endhost/assetcontract/{\ding{186} Redeem}/1.5/0.9,
			3.6/{->,pos=0.35}/assetcontract/as/{\ding{187}}/1/0.25,
			4.2/{<-,pos=0.41}/assetcontract/as/{\ding{189} Deliver Res.}/1/1.21,
			4.2/{<-,pos=0.5}/endhost/assetcontract/{\ding{189} Reservation}/1.5/2
		} {
			\draw[line width=3,sloped, white] ($(\l)-(-\spacing,\i*\dist)$) -- ($(\r) -(\spacing,\i*\dist + \o*\slope*\dist)$);
			\draw[thick,sloped,apply style/.expand once=\s] ($(\l)-(-\spacing,\i*\dist)$) -- node (label\thenodeidx)[above,text width=0.5*\v*\hdist,align=center,fill=white,inner sep=1]{\scriptsize\t} ($(\r) -(\spacing,\i*\dist + \o*\slope*\dist)$);	
			\stepcounter{nodeidx};
		}
		
		 \node (ingress1) [smallasset, fill=redfte, anchor=east] at ($(endhost)+(-0.1, -1*\dist)$) {};
		 \node[left=0cm of ingress1,align=center,inner sep=0] {\scriptsize\ding{182}};
		 
		 \node [largeasset, fill=bluefte, above right=0.1cm and 0cm of label1.east, anchor=west] {};
		 \node  [largeasset, fill=bluefte, anchor=west] at ($(as)+(0.1, -1.85*\dist)$) {};
		 \node  [largeasset, fill=bluefte, above right=0.1 and 0 of label2.east, anchor=west] {};
		 
		 \node (egress1) [largeasset, fill=bluefte] at ($(market)+(0.35, -2.75*\dist)$) {};
		 \node [smallasset,color=white, fill=white, anchor=north west, draw=white, line width=1pt] at ($(egress1.north west)+(-1pt,1pt)$) {};
		 \node [smallasset,color=bluefte, fill=bluefte, anchor=north west] at ($(egress1.north west) + (-2pt, 2pt)$) {};
		  
		  \node [smallasset, fill=bluefte, anchor=west,right=0.0 of label5.east] {};
		  
		  \node (egress2) [smallasset, fill=bluefte, anchor=east] at ($(endhost)+(-0.1, -3.35*\dist)$) {};
		  \node [smallasset, fill=redfte,left=0.0 of egress2] {};
		  
		  \node (ingress2) [smallasset, fill=redfte, right=0.0 of label6.east, anchor=west] {};
		  \node [smallasset, fill=bluefte,right=0.0cm of ingress2] {};
		  
		  \node (ingress3) [smallasset, fill=redfte, above right=1.5pt and 0.0 of label7.east, anchor=west] {};
		  \node (egress3) [smallasset, fill=bluefte,right=0.0cm of ingress3] {};
		  \node[rectangle, draw=black, fit=(ingress3) (egress3), inner sep=1pt] {};		  
		  
		  \node (ingress4) [smallasset, fill=redfte, anchor=west] at ($(as)+(0.15, -3.9*\dist)$) {};
		  \node (egress4) [smallasset, fill=bluefte,right=0.0cm of ingress4] {};
		  \node[rectangle, draw=black, fit=(ingress4) (egress4), inner sep=1pt] (redeemreq1){};		
		  \draw[thick,->] ($(as)-(-0.15,3.6*\dist)$) arc [
		  	start angle=135, delta angle=-270,radius=0.35
		  ]; 
		  \node [inner sep=1, above=0.2cm of redeemreq1]{\scriptsize\ding{188}};	
		  \node (ingress5) [smallasset, fill=redfte, above right=2pt and 2pt of label8.east, anchor=west] {};
		  \node (egress5) [smallasset, fill=bluefte,right=0.0cm of ingress5] {};
		  \node[rectangle, draw=black, fit=(ingress5) (egress5), inner sep=1pt] {};	
		\end{tikzpicture}
}
	\caption{Control-plane interactions between the asset and market smart contracts, an AS, and an end host. The figure shows how assets are issued (\ding{183}) and listed on a marketplace (\ding{184}), how an end host obtains an egress asset (\ding{185}) that matches a previously obtained ingress asset (\ding{182}), and how the reservation is redeemed (\ding{186}--\ding{189}).}
	\label{fig:controlplane}
\end{figure}

In the following, we will first describe the structure of a bandwidth asset, and then explain the registration process for ASes, before we detail how an end host obtains a reservation.
\Cref{fig:controlplane} illustrates this process in an example, in which we assume that the AS has previously registered with the asset contract, and that the end host already owns an ingress asset for the desired flyover reservation (\ding{182}), e.g., from a previous purchase.

\para{Asset Representation}
As reservations should be possible for both long and short time periods as well as for large and small bandwidth allocations, issuing an individual asset for each possible reservation quickly becomes infeasible.
Because of this, we represent reservations using assets that are splittable in both the time and bandwidth dimension, subject to the limitations of the provided granularities (see below).
This allows an AS to publish one large bandwidth asset, which can then be split up into smaller bandwidth assets as needed, which will be explained below.

Additionally, providing individual assets for all possible pairs of ingress/egress interfaces can quickly lead to a space explosion.
We avoid this by making each asset represent a reservation on a single interface used either as ingress or egress for the reservation.
To later make use of the reservation, an asset for both an ingress and egress interface from the same AS representing the same bandwidth and time intervals are needed.

Each bandwidth asset contains the following attributes:
\begin{description}
	\item[AS Identifier:] The identifier of the AS offering the reservation.
		This is set automatically after authenticating the AS.
	\item[Bandwidth:] The bandwidth of the reservation. Corresponds to $\bw$ on the data plane.
	\item[Start Time:] The start time of the reservation represented by the asset. Corresponds to $\strt$ on the data plane.
	\item[Expiration Time:] The expiration time of the reservation represented by the asset. Corresponds to $\strt + \duration$ on the data plane.
	\item[Interface:] The ID of the AS interface for which the reservation is valid. Corresponds to $\In$ or $\Eg$ on the data plane.
	\item[Ingress/Egress:] An indicator describing whether the reservation allows the use of the interface as ingress or egress.
	\item[Time Granularity:] The minimum duration for which the AS wants to support reservations. Supporting very short reservations may not be in the interest of every AS, thus we allow each AS to set this threshold.
		Each asset can be split in the time dimension as long as the duration of each new asset is a multiple of this granularity.
	\item[Minimum Bandwidth:] The minimum bandwidth of reservations.
		This parameter allows an AS to limit the number of concurrent reservations, see also \cref{sec:policing}.
		Each asset can be split in the bandwidth dimension as long as the bandwidth of each new asset exceeds this minimum.
\end{description}

\para{AS Registration}
Our asset smart contract requires a registration process, which ensures that the entity issuing an asset is actually authorized to do so; i.e., an entity cannot simply create an on-chain asset that represents a reservation at some AS that is not under their control.
Since we assume that a PKI for ASes is in place, each AS could simply sign each asset that they issue, which would allow any user to verify the validity of an asset before obtaining it.

This naive approach can be improved to only require authentication once, during an AS registration process, in which the AS provides its AS certificate to the asset smart contract and proves that it is in possession of the corresponding private key.
At this point, the smart contract issues an authorization token to the AS that contains the AS identifier.
The asset contract then enforces that (i) only the owner of such an authorization token, i.e., registered ASes, can issue bandwidth assets, and (ii) that the AS identifier listed in each issued asset corresponds to the AS identifier contained in the authorization token that is provided during issuance.

\para{Issuance} Each registered AS can issue bandwidth assets for ingress and egress interfaces by calling the \emph{issue} function of the asset contract, specifying the attributes of the asset.
In Step \ding{183} in \cref{fig:controlplane}, the AS issues a large bandwidth asset for an egress interface, which the contract immediately transfers to the AS.
At this point, the asset is owned by the AS and can be transferred arbitrarily.
In our example, the AS now lists this asset for sale by sending it to a market contract (\ding{184}), which allows the market contract to modify the asset, governed by the restrictions put in both the market and asset contracts, e.g., dictating how assets can be split.

\para{Asset Splitting \& Purchase}
In our example, the AS has issued, and listed for sale, an asset that represents a longer time interval and larger bandwidth than the end host wants to purchase, namely, an asset that matches the dimensions of the previously obtained ingress asset (\ding{182}).

Luckily, the owner of a bandwidth asset---which, in the example is the market contract---can split it in the time or bandwidth dimension if the split conforms to the restrictions on the time granularity and minimum bandwidth, resulting in two ``smaller'' assets.
When splitting an asset, the attributes Start/Expiration Time and Bandwidth are adjusted according to the split, whereas the other attributes remain unchanged.
To split an asset into multiple chunks or in both dimensions, the process is repeated.

Thus, in our example (\cref{fig:controlplane}), the end host can buy a fraction of the asset by sending the required payment to the market contract and specifying the desired time interval and bandwidth (\ding{185}).
The payment is forwarded to the seller (in this case the AS) and the asset that was split off is transferred to the end host,
who now becomes its owner. This allows them to transfer or resell the asset, to split it again, or to fuse it with other compatible assets.



\para{Asset Redemption}
Before a reservation can be used on the data plane, a pair of ingress/egress assets that represent this reservation need to be \emph{redeemed}.
In this step, the owner of said assets exchanges them for the required data-plane information.
This is necessary since ownership of assets themselves cannot directly allow use of the network resource, as this requires information (in particular the authentication key $A_K$) that (i) needs to remain secret, (ii) depends on attributes that change with each split of the asset, and (iii) depends on both assets that are combined for the reservation.

Once an end host has obtained compatible ingress and egress assets (i.e., same AS, validity period, bandwidth) and wants to use the reservation, the end host---as owner of the assets---sends both assets, as well as 
an ephemeral public key
to the asset contract to redeem them (\ding{186}).
The contract wraps the assets and the public key into a \emph{redeem request} (represented as another on-chain asset) which is then transferred to the issuing AS (\ding{187}).
Once the AS receives this redeem request, it creates the information required for the reservation (i.e., $\resinfo_K, A_K$) as described in \cref{sec:details:authentication} and encrypts it with the public key from the redeem request (\ding{188}).

In this step, the AS also assigns the reservation's $\resid$.
To keep monitoring as efficient as possible (see \cref{sec:policing}), per-ingress unique IDs are preferred; however, an AS is free to assign IDs differently as long as it can uniquely identify and monitor a reservation based only on the \resinfo{}.

The AS then sends the encrypted reservation (together with the redeem request asset) back to the asset contract, from which it is received by the client (\ding{189}). The contract additionally destroys the bandwidth assets, which can thus no longer be traded.
Finally, the client decrypts the received information and forwards $\resinfo_K$ and $A_K$ to its \name-enabled local applications to use on the data plane.

\para{Atomic End-to-End Guarantees}
One crucial property we get from using blockchain-based assets is that multiple transactions can be made atomic, i.e., they either all succeed or all fail.
Thus, even though each asset represents only an interface of a hop on the path, a source can create an atomic transaction to buy reservations for all hops on a path:
If it succeeds, the source receives an end-to-end bandwidth reservation; if it fails, no money is lost (except minimal transaction fees).

The atomicity of blockchain transactions solves one of the most difficult tasks in hop-based bandwidth-reservation systems:
Obtaining end-to-end reservations without active coordination between the involved ASes and with no significant partial-failure costs.

\subsection{Data Plane}\label{sec:details:dataplane}

The data plane for \name flyovers is lightweight, allows stateless traffic authentication on routers, and only requires a small state for policing (see \cref{sec:policing}).

When sending a packet over the reservation, after the start time has passed, the source authenticates the packet with the \emph{packet authentication tag}
\begin{equation}
	V_K= \prf{A_K}{\dst, \len(\pkt), \ts}[:\ell_{\tg}],\label{eq:tag}
\end{equation}
where $\len(\pkt)$ is the length of the packet (including header) and $\ts$ is a high-resolution timestamp that is unique for each packet, used for traffic policing.
$\dst$ is the destination address, which is included to mitigate reservation stealing attacks (see \cref{sec:security}) and $[:\ell_{\tg}]$ denotes truncation to $\ell_{\tg}$ bytes, which is a parameter configured at the protocol level.
Truncating the tags reduces the effort required for brute-force attacks.
However, since the tags are only valid for a short amount of time (see below), and brute-forcing them requires an online attack, this is sufficient for our use case (see \cref{sec:security}) and allows saving valuable space in the packet headers. In systems like SCION~\cite{chuat2022complete}, where each hop already contains a \emph{Hop Field MAC} to authenticate the forwarding information, $V_K$ can be aggregated with this Hop Field MAC into an aggregate MAC~\cite{katz2008aggregate} by XORing the two, which saves additional space (see \cref{sec:headerspec:flyover}).

The packet is then transmitted, containing the reservation information \eqref{eq:resinfo} and tag \eqref{eq:tag} for each hop with a reservation:\footnote{Parts of the \resinfo{} are already contained in the SCION standard header.}
\begin{equation}
	(\ts, \len(\pkt), \resinfo_1, V_1, \dots , \resinfo_K, V_K, \dots).
\end{equation}

Upon receiving a packet, each border router performs the following steps:
\begin{enumerate}
	\item Check if the packet contains a reservation.
	      If not, treat it as best-effort traffic.
	\item Verify the authenticity of the reservation information in the packet to ensure that the packet is using bandwidth of a valid reservation.
	      The border router only needs to recompute $A_K$ and $V_K$, and compare the latter to the one in the packet to verify the validity of the reservation.
	      If the authentication fails, the packet is dropped.
	\item Check if the timestamp $\ts$ in the packet is recent---i.e., between $t - \delta - \Delta$ and $t+\delta$, where $t$ is the current time, $\delta$ is the maximum clock skew, and $\Delta$ is the maximum packet age---and that it falls within the validity period of the reservation.
	      If this is not the case, treat the packet as best-effort traffic.
	      Otherwise continue with the next step.
	\item Monitor the reservation traffic, see \cref{sec:policing}.
	\item Forward the packet with priority if there is still sufficient bandwidth available in the reservation, otherwise forward with best effort.
	      Packets are not dropped to ensure that benign bursts of traffic---which can occur through network elements that are not under the control of the source---do not degrade performance to below best-effort performance.
\end{enumerate}

Since reservation traffic is prioritized over best-effort traffic, it will be able to pass, even if the network is congested.
However, unused bandwidth from a reservation is not wasted, since it is still available for best-effort traffic.

\subsection{Traffic Policing}
\label{sec:policing}

\name is designed to allow every AS to perform deterministic monitoring and policing for its own reservations to ensure that it can provide its guarantees. This is in contrast to systems like Colibri~\cite{giuliari2021colibri}, where the source AS performs deterministic traffic policing and on-path ASes only perform probabilistic monitoring.

For each reservation, policing can be implemented with a simple token-bucket algorithm using a single \SI{8}{\byte} counter, as described below.
Since an assigned $\resid$ is unique for the particular ingress interface during the reservation's validity period, an AS can simply keep an array of $\fl{ResID}_\mathrm{max}$ counters, where $\fl{ResID}_\mathrm{max}$ is the highest assigned \resid{} of all its reservations, and use the \resid{} in the packet header as an index.
If the \resid{}s are assigned such that $\fl{ResID}_\mathrm{max}$ is sufficiently small, the policing array can fit into a processor's cache, making policing very efficient.

\para{ResID Assignment} Assigning \resid{}s is an instance of the online interval coloring problem which can be solved efficiently~\cite{gyarfas1988line,kierstead1981extremal,kashyop2020dynamic,kierstead2016first}. However, online graph coloring algorithms cannot assign the optimal coloring and instead, in the worst case, assign a maximum color $R\cdot \chi$, where $\chi$ is the chromatic number of the graph and $R$ is called the \emph{competitiveness}  of the coloring algorithm.
In our case, the competitiveness is the ratio between the largest assigned \resid{} and the maximum number of concurrent reservations (i.e., $\fl{TotalBW} / \fl{MinBW}$).

By setting the total bandwidth $\fl{TotalBW}$ available for reservations during a time period as well as the minimum bandwidth $\fl{MinBW}$ for reservations, an AS can ensure that the possible values for the \resid{}s are between 0 and $\fl{ResID}_\mathrm{max} := R\cdot\fl{TotalBW} / \fl{MinBW}$ in the worst case.
An AS can always ensure that this array fits into the 
processor's cache by choosing appropriate values for \fl{TotalBW} and \fl{MinBW}.  
Consider the following two examples (using $R=3$):\footnote{In the optimal online algorithm $R = 3$~\cite{kierstead1981extremal}, although other algorithms perform better in practice for most interval graphs~\cite{gyarfas1988line}.}
\begin{enumerate}
	\item An ingress interface with a total bandwidth of \SI{100}{Gbps} (a common link size in the current Internet~\cite{peeringdb}) and a min bandwidth of \SI{100}{kbps} (sufficient for VoIP~\cite{zoomrequirements}):
	This results in $\fl{ResID}_\mathrm{max} = \num{3e6}$ and a policing array of \SI{24}{\mega\byte}, which fits into the L3 cache of many current processors.
	
	\item The same total bandwidth but a minimum of \SI{4}{Mbps}, which is sufficient for 1080p video calls~\cite{zoomrequirements}:
	This results in $\fl{ResID}_\mathrm{max} = \num{75000}$ and a \SI{600}{\kilo\byte} policing array, which fits into the L2 cache of current processors.
\end{enumerate}

Using the \resid{} directly as an index into the policing array and keeping the reservation bandwidth in the packet header makes the policing algorithm extremely simple.

\newcommand{\bt}{\fl{BurstTime}}
\para{Traffic Policing Algorithm} \Cref{alg:policing} describes a simple and highly efficient algorithm, adapted from the algorithm presented by Wyss et al.~\cite{wyss2022protecting}, for deterministic policing, which only keeps an 8-byte timestamp per reservation in an array plus a global parameter \bt{}.

This parameter limits the allowed ``burstiness'':
intuitively, a sender cannot send more than \bt{} worth of traffic compressed into a single burst.
It should be set sufficiently small such that the router's buffers can absorb some synchronized bursts and to limit jitter, but large enough that it does not cause unnecessary constraints on packet sizes for small reservations.

Given current trends in router buffers~\cite{sharada2023sizing}, a value of roughly \SI{50}{\ms} seems reasonable.
Note that for very small reservations (below \SI{240}{kbps}), this limits the maximum packet size to below \SI{1500}{\byte}.
However, specifically for VoIP applications, this is not an issue, as those anyway send a packet roughly every \SI{20}{\ms}~\cite{ciscovoip} and thus do not trigger the burst prevention.

Processing a packet requires very few operations, namely i) getting the current time, ii) 1 load operation, iii) 2 comparisons and branches, iv) 1 division\footnote{If desired, the division in line~\ref{alg:policing:div} of \cref{alg:policing} can be replaced by a load and multiplication by storing the inverse of all possible 1024 \bw{} values in an array (assuming a 10-bit field as described in \cref{sec:headerspec:flyover}).}, v) 2 additions, and vi) 1 store operation (in the successful case).

\begin{algorithm}[h]
	\caption{
		Simple policing algorithm for \name which assumes unique values for the \resid of all reservations on a given ingress interface.
		This algorithm locally stores an array of timestamps (one for each possible value of \resid) and a global parameter \bt{}.
		All other inputs (\resid, \bw{}, \fl{PktLen}) are part of the authenticated packet information.
		The algorithm is a slight adaptation of the algorithm presented in Appendix~E of Wyss et al.~\cite{wyss2022protecting}.
	}
	\label{alg:policing}
	\begin{algorithmic}[1]
		\renewcommand{\algorithmicrequire}{\textbf{Input:}}
		\Require \fl{TSArray}, \bt{} (local store)
		\Require \resid, \bw{}, \fl{PktLen} (packet)
		\Ensure fwd\_type
		
		\Function{BandwidthMonitoring}{\resid, \bw{}, \fl{PktLen}}
		\State $\fl{now} \gets \mathrm{time}()$
		\State $\fl{TS} \gets \max\mathopen(\fl{TSArray}[\resid], \fl{now}) + \fl{PktLen} / \bw{}$ \label{alg:policing:div}
		\If{$\fl{TS} \leq \fl{now} + \bt{}$}
		\State $\fl{TSArray}[\resid] \gets \fl{TS}$
		\State \Return \fwdfly{}
		\Else
		\State \Return \fwdbest{}
		\EndIf
		\EndFunction
	\end{algorithmic}
\end{algorithm}


\section{Security Analysis}\label{sec:security}

In this section, we analyze the security of \name. We first introduce the adversary model, then describe the security properties, and finally analyze the security of the control and data planes.

\subsection{Adversary Model}\label{sec:adversarymodel}

\begin{figure}[t]
	\centering
	\centering
\begin{tikzpicture}[
    asnode/.style={draw=black, fill=white, circle, minimum size=0.8cm, line width=2pt, inner sep=0pt, outer sep=0pt},
    aslink/.style={draw, line width=2pt, black},
    cparrow/.style={-stealth, redfte, line width=2pt},
    doublecparrow/.style={stealth-stealth, redfte, line width=2pt},
    numbernode/.style={black, minimum size = 0pt, outer sep=0pt, inner sep=0pt,font={\large}},
    csbox/.style={line width=1.5pt,anchor=center,draw,minimum size=0.6cm,inner sep=0pt, rounded corners=0.2cm, fill=white, font={\bfseries}},
    segr/.style={fill=redfte!50, draw=redfte, line width=2pt},
    ]

    \node [asnode] (ASs) at (-5,0) {};
    \node [asnode] (ASx) [above right= 0.3cm and 0.7cm of ASs] {};
    \node [asnode, draw=redfte] (ASy) [below right= 0.3cm and 0.7cm of ASs] {};
    \node [asnode] (ASw) [below right = 0.3cm and 0.7cm of ASx] {};
    \node [asnode] (ASz) [right = 1.4cm of ASw] {};
    \node [asnode] (ASd) [right = 1.4cm of ASz] {};

    \node (HS) [numbernode, fill=none] at (ASs.west) {\laptop{}};


    \draw [aslink] (ASs) -- (ASx);
    \draw [aslink] (ASs) -- (ASy);
    \draw [aslink] (ASy) -- (ASw);
    \draw [aslink] (ASx) -- (ASw);
    \draw [aslink] (ASw) -- (ASz);
    \draw [aslink] (ASz) -- (ASd);

    \draw [segr] ([xshift=-10pt, yshift=8pt] ASz.west) rectangle ([xshift=10pt, yshift=-8pt] ASz.east);
    \draw [segr] ([xshift=-10pt] ASz.west) ellipse (3pt and 8pt);
    \draw [segr] ([xshift=10pt] ASz.east) ellipse (3pt and 8pt);

    \draw [bluefte, smooth, line width=3pt, line cap=round] plot coordinates {(HS.north east) ([yshift=3pt]ASx) ([yshift=6pt,xshift=5pt] ASw) ([yshift=4pt,xshift=-20] ASz)};
    \draw [greenfte, line cap=round, dash pattern={on 7pt off 7pt}, smooth, line width=3pt] plot coordinates {(HS.south east) ([yshift=-3pt]ASy) ([yshift=-6pt,xshift=5pt] ASw) ([yshift=-4pt,xshift=-20] ASz)};

    \draw [bluefte, smooth, line width=3pt, -stealth, line cap=round] plot coordinates {([yshift=4pt,xshift=20] ASz) ([yshift=4pt] ASd)};
    \draw [greenfte, line cap=round, dash pattern={on 7pt off 7pt}, smooth, line width=3pt, -stealth] plot coordinates {([yshift=-4pt,xshift=20] ASz) ([yshift=-4pt] ASd)};

    \node (L1) [right, align=left] at ([xshift=15pt,yshift=-15pt]ASy.east) {Adversary (not on $P$)\\can observe the shared reservation};
    \node [pin={[pin distance=1cm,align=left]90:{Target path $P$}}] at (ASw.north east) {};
    \node [pin={[pin distance=0.4cm,align=left]-90:{Path $Q$}}] at ([xshift=-0.4cm, yshift=0.1cm]ASy.west) {};
    \node [pin={[pin distance=0.4cm,align=left]90:{Target AS $T$}}] at (ASz.north) {};
    \draw [very thin, gray] (ASy.south east) -- (L1.west);

\end{tikzpicture}

	\caption{On-reservation-set adversary. The source is using two paths $P$ (solid line) and $Q$ (dashed line) concurrently, with a single shared reservation on AS $T$. Despite not being on the target path $P$, the adversary can observe authentication tags that are also valid for path $P$, since the same reservation is used for path $Q$.}
	\label{fig:on-reservation-set}
\end{figure}
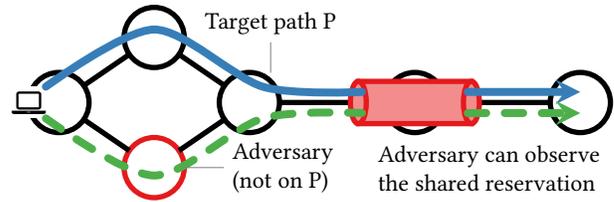

To analyze \name's security, we consider an adversary that controls a set of hosts, ASes, and/or routing elements.
The adversary is distributed and can drop, modify, duplicate, inject, or delay traffic but not break cryptography.
We further make the following assumptions to allow a meaningful analysis of the system's properties:
\begin{enumerate}
	\item There exists a routable path of honest ASes from source to destination.
	      Otherwise, no availability guarantees can be provided in any system since the adversary can always drop the traffic and prevent any communication.
	\item Every honest source and AS is able to communicate with the control-plane elements  to obtain reservations.
	      For our control plane, this means that each source and each AS must have access to a path of honest ASes that connects them to the blockchain network on which the control-plane smart contract is hosted.
	      In other words, we assume that the adversary cannot eclipse~\cite{heilman2015eclipse,wuest2016ethereum} an honest party.
	\item There is an efficient (in the economic sense) market that allows bandwidth-reservation assets to be priced correctly.
\end{enumerate}

We distinguish several different adversaries.
Considering a source $S$ and a destination $D$ communicating over a path $P$, we distinguish the following adversary types:
\begin{description}
	\item[Compromised source:]
		The adversary tries to gain an unfair advantage as the source of the communication.
	\item[On-path adversary:]
		The adversary controls at least one AS or routing element on the path~$P$ (but not the source~$S$).
	\item[On-reservation-set adversary:]
		This type of adversary is illustrated in \cref{fig:on-reservation-set}, and can only exist if the source simultaneously uses two paths $P$ and $Q$ that share a reservation on at least one target hop.
		The adversary then controls at least one AS that is on path $Q$ but not on $P$ and targets reservation traffic on path $P$.
		Thus, this adversary can observe valid packet authentication tags for the reserved downstream ASes that are shared between paths $Q$ and $P$.
	\item[Off-path adversary:]
		The adversary controls a set of hosts and ASes, but is not one of the previously defined types.
\end{description}

\subsection{Security Properties}\label{sec:securityproperties}

\name provides the following security properties, on the control plane (\labelcref{prop:secure}, \labelcref{prop:fair}) and data plane (\labelcref{prop:overuse}, \labelcref{prop:qos}):
%
\begin{enumerate}[label=\textbf{C\arabic*},ref=C\arabic*]
	\item\label{prop:secure} \emph{Secure reservation establishment.}
	The control plane allows an honest host to securely establish a reservation, i.e., the control-plane guarantees confidentiality for the reservation authentication keys, and integrity of the reservation information.
	In addition, the adversary cannot perform \emph{denial-of-capability attacks} that prevent an honest host from obtaining reservations on an honest AS hop---provided that there is sufficient bandwidth available.
	\item\label{prop:fair} \emph{Fair bandwidth distribution.}
	The adversary cannot gain more than a fair share of the bandwidth available for reservations, by, e.g., creating multiple accounts or using large numbers of hosts (Sybil attack).
	This abstract property depends on the specific notion of fairness.
	In this paper, we consider the notion of \emph{economic fairness}: An adversary cannot obtain a larger reservation than other nodes without paying the fair market price for it. 
	Equivalently, the adversary cannot starve honest nodes for bandwidth without paying the respective price.
\end{enumerate}
%
\begin{enumerate}[label=\textbf{D\arabic*},ref=D\arabic*]
	\item\label{prop:overuse} \emph{Overuse protection.} An adversary cannot overuse or spoof a reservation. The adversary cannot use a forged reservation, or a valid reservation (at an honest AS hop) outside its limits, i.e., before start time, after end time, or for more bandwidth than granted.
	\item\label{prop:qos} \emph{Quality of Service (QoS).}
	Legitimate hosts should benefit from the QoS guarantees promised by reservations at an honest AS. These guarantees include the amount of bandwidth reserved, but may also incorporate, e.g., bounds on jitter, latency, or other network-layer properties that may be provided by the AS inside their domain.
	As an on-path adversary can always perform a DoS attack, and thus disrupt QoS, by simply dropping packets (as in all bandwidth-reser\-vation systems), we do not consider this adversary type in our analysis of this property.
\end{enumerate}

\subsection{Control-Plane Analysis}

\para{Secure reservation establishment (\labelcref{prop:secure})}
Secure reservation establishment is ensured through our assumptions on the control plane in combination with the registration of ASes and the encryption of $\resinfo_K, A_K$ during redemption.

Since we assume that a PKI is in place for ASes (see \cref{sec:overview:assumptions}) and ASes prove to the asset smart contract that they are in possession of the secret key corresponding to their certificate, each issued asset provides authenticity.
Due to the access-control guarantees provided by the blockchain and the smart contracts (only the owner of an asset can use the asset in a transaction), no other party can tamper with the message sent to deliver the reservation, which ensures its integrity.
Confidentiality is provided by encrypting the asset with the public key of the asset owner (i.e., the host). This public key is sent to the AS together with the asset through the smart contract, which guarantees authenticity of the key (again, since only the owner can use the asset) and therefore confidentiality of messages encrypted with this key.

Finally, our assumption that honest hosts and ASes are not eclipsed (\cref{sec:adversarymodel}) and can therefore always reach the blockchain hosting the smart contracts ensures the availability of the marketplace, and protects against attacks that prevent an honest host from obtaining a reservation.

\para{Fair bandwidth distribution (\labelcref{prop:fair})}
Fairness and Sybil resistance are guaranteed by allowing the assets to be freely tradeable and our assumption that an economically efficient market exists that enables price discovery for a fair market value.
The existence of such a market then guarantees \emph{economic fairness}. Even if an adversary was able to obtain reservations at a price that is lower than the value of the reservations at the time of an attack (e.g., by buying the reservations far in advance when they are valued lower), they would suffer the opportunity cost of not reselling the bandwidth assets.
Thus, the cost of the attack is always at least equal to the value of the corresponding reservation.
The same applies to starving honest sources of bandwidth:
To do so, the adversary would have to reserve the full hop bandwidth themselves---and therefore pay for it---becoming its legitimate owner.
Note that the adversary gains nothing from creating multiple hosts or accounts, i.e., a Sybil attack, since they need to pay the market value for each asset, independent of the number of accounts that they are using to obtain reservations.

A concrete market design or analysis thereof is out-of-scope for this work. However, there are several well-known market mechanisms that could plausibly be deployed to achieve efficient price discovery and allocation of bandwidth reservations. One possibility are Vickrey-Clarke-Groves (VCG) auctions~\cite{vickrey1961counterspeculation,clarke1971multipart,groves1973incentives}, in which participants bid on multiple items simultaneously and the final allocation maximizes the total utility. These auctions are provably strategy-proof and economically efficient under a broad set of assumptions and could be utilized to vote for bandwidth reservations along a path. However, this would require additional rounds of communication with a smart contract as well as discrete rounds in which the auctions complete.

Alternatively, a more lightweight approach could involve a spot market with posted prices, similar to our prototype, where Autonomous Systems (ASes) list bandwidth offers in real time, and users select the best-priced paths that satisfy their requirements. Such markets offer responsiveness and deployability, albeit with potential inefficiencies under high volatility or demand shocks. Hybrid mechanisms, e.g., combining advance reservation auctions with short-term spot market resales, could help stabilize pricing while maintaining flexibility for bursty traffic.

Crucially, the path diversity in the underlying SCION architecture enhances the viability and liquidity of any such market. Between most source-destination pairs, there are multiple disjoint or partially overlapping paths, recent measurements in SCIERA, an educational network that is part of the production SCION network, show that between most source/destination pairs, there are more than twenty, and up to one hundred, paths available, and at least two paths exist between any pair~\cite{wirz2025sciera}. This diversity provides many alternative paths that constitute substitutable goods in economic terms: if one path segment is overpriced or congested, alternative paths can be used instead. This not only reduces monopolistic pricing risks, but also improves price stability and makes the market more resilient to manipulation or scarcity-induced spikes.

We argue that the design of \name enables a range of plausible mechanisms that are capable of achieving sufficient price discovery and economic fairness to support our assumption. Future work may explore these mechanisms in more detail, potentially adapting tools from existing pricing research and auction theory to the unique constraints of inter-domain bandwidth reservations.

\subsection{Data-Plane Analysis}\label{sec:security:data}

\para{Overuse protection (\labelcref{prop:overuse})}
\name protects against reservation overuse and spoofing through its per-packet reservation authentication and traffic policing.
The authentication tag is computed over the full reservation description and a secret value only known to the granting AS, and each AS performs traffic policing based on these protected values.
Therefore, an adversary that either tries to spoof a reservation or to overuse a valid reservation must be able to forge an authentication tag.
Since we assume that the authentication tags are computed using a secure PRF, the authentication tags are a secure MAC with security parameter $\ell_{\tg}$.

This ensures that the only possible attack vector is a brute-force attack on the authentication tag, the feasibility of which depends on their length $\ell_\mathrm{tag}$.
To save space in the packet header, tags are ideally as short as possible, while providing sufficient security in practice.

A brute-force attack on \name cannot be performed offline, as it is not possible to check the validity of a tag without knowledge of the key.
This means that for each candidate tag, a packet needs to be sent to the reservation AS, which forces the adversary to brute-force the tag in an online attack.
Brute-forcing a valid tag only allows an adversary to use it for a short time, namely the validity period (e.g., 1 second) of a packet with the provided timestamp.
Therefore, the tag length $\ell_\mathrm{tag}$ only needs to be chosen such that an online attack becomes infeasible (or prohibitively expensive) for an adversary.
In our implementation, we use a tag length of 6 bytes, which would require the adversary to send (on average) more than 140 trillion packets to have one success, which is prohibitively expensive in practice.
Note that brute-forcing the key used for the PRF---which would allow sending packets until the expiration time of the brute forced reservation---depends on the key length, not the tag length, and is therefore infeasible.

\para{Quality of Service (\labelcref{prop:qos})}
In an honest AS, QoS is provided by prioritizing the reservation over best-effort traffic, ensuring that the source can use the full reservation bandwidth.

An adversary can try to cause delayed or dropped packets, e.g., by introducing congestion at the reserved hop.
Since an honest AS does not hand out more reservations than it can handle and prioritizes reservation traffic, congestion caused by best-effort traffic does not impact reservations.

This does not exclude the threat of other DoS attacks, which we analyze separately for off-path and on-reservation-set adversaries.
As discussed above, brute-force attacks on authentication tags are not feasible in practice.
Therefore, off-path adversaries cannot forge authentication tags for an honest AS hop, which prevents them from disturbing reservation traffic, since they can only either send best-effort traffic, or traffic belonging to a valid reservation.

Since reservations are granted for individual hops, a host may use the reservation for a specific hop on multiple paths.
In this case, an on-reservation-set adversary could observe valid authentication tags for reservations on hops that are used by the source for multiple paths.
This enables such an adversary to perform a DoS attack (\ref{prop:qos}) on these hops.

Namely, an on-reservation-set adversary can observe authentication tags for the reserved hop from packets that were sent on a path that includes the adversary.
The adversary can then fabricate packets that will successfully pass authentication at the reserved hop by duplicating the observed packet authentication tags (optionally dropping the original packets) which is possible during the validity period of the observed packet (e.g., 1 second).
Since \name does not require ASes to deploy duplicate suppression, this can be used for a DoS attack on packets that do not flow through the attacker: by sending large amounts of traffic on the reservation to exhaust the reservation's bandwidth limit at a downstream AS, causing legitimate traffic arriving on a different path at the downstream AS to get dropped by the policing system.

To prevent such attacks, it is sufficient to make a separate reservation for shared hops on each path and thus use each reservation exclusively on one path at a time.
Whether to use a single reservation on multiple paths or to obtain separate reservations for each path is a trade-off between convenience and security.
The decision between these options can be made by the source, depending on whether they trust ASes on all paths that they use, or only the ASes on the current path.

\para{Reservation-stealing as incentive} One possible motive for an attacker is \emph{reservation-stealing}.
An on-path or on-reservation-set adversary can attempt to use a reservation from another party for themselves by reusing the authentication tag of a valid packet and replacing the payload.
The effect of such an attack for the source is the same as any other DoS attack, and our analysis above applies equally.
However, reservation-stealing provides an additional incentive, as the compromised AS can benefit from using the reservation.

As a mitigation, the computation of the authentication tag includes the destination address. This ensures that an adversary can only benefit if they either (i) send traffic to the same destination, or (ii) control an additional on-path AS after the reservation that redirects the packet to the new destination.


\para{On the Absence of Duplicate Suppression and Gateways}
\label{sec:nodupsupp}

Previous bandwidth-reservation designs (Helia~\cite{wyss2022protecting}, Colibri~\cite{giuliari2021colibri}) require duplicate suppression at on-path ASes and gateways at the source AS.
We analyze here why \name can work without these cumbersome subsystems while still achieving similar security and availability properties.

Both duplicate suppression and the gateway are in place mainly to prevent framing attacks, in which the adversary sends a large amount of seemingly legitimate traffic to frame the source AS, and cause repercussions for the source, such as precluding new reservations.
These attacks may work in two ways:
\begin{itemize}
	\item An on-path adversary may duplicate packets and send them at a high rate.
	These packets will pass authentication, and it is therefore necessary to remove them before they are accounted for in the monitoring.
	\item A compromised or malicious host who has access to hop authentication keys may relay them outside the AS, allowing an off-path adversary to generate seemingly legitimate traffic.
	To prevent this, the gateway was introduced such that authentication keys are not accessible to endhosts.
	
	It is important to note that this problem stems from the fact that previous systems rely on ASes to manage reservations for scalability reasons, i.e., to aggregate the control plane and monitoring tasks at the AS level.
\end{itemize}

In our system, these cases are not a problem for two reasons:
\begin{enumerate}
	\item There is no penalty for overusing a reservation:
	in case of overuse, packets are simply dropped.
	Therefore, framing attacks are never successful.
	\item The scalability of the system does not rely on AS-level aggregation.
	Therefore, any entity can obtain authentication keys without external mediation from their AS.
	The gateway is therefore not strictly necessary, as the monitoring is directly imputed to the source host and the AS is not directly responsible.
\end{enumerate}

The only type of attack that would be prevented in \name by introducing duplicate suppression are reservation-DoS attacks from on-reservation-set adversaries.
We believe, however, that the practical advantages of removing duplicate suppression from the system requirements outweigh this downside, in particular, since the same attacks can be prevented by obtaining separate reservations for each used path.

Further, duplicate suppression can be incrementally added (the system is designed to include unique packet IDs that can be used for duplicate suppression) by individual ASes.
Note that, for on-path adversaries, the lack of duplicate suppression does not introduce new attack vectors, since an on-path adversary is always able to block traffic from the source.

The gateway's function of enhancing the system's scalability---by multiplexing multiple AS-internal reservations in one single inter-domain bandwidth reservation---is still beneficial, and our system readily supports the implementation of gateways to this end.
However, the gateway is not \emph{required}, which makes \name more flexible and improves the speed at which it can be implemented and deployed in practice.


\section{Control-Plane Evaluation}
\label{sec:evaluation_control}

In this section, we describe the implementation of our control plane and evaluate its performance. Additional evaluation results are provided in \cref{app:additional}.
The most relevant analysis here is the cost and time required to make a reservation on the marketplace. Note that the scalability of \name's control-plane operations is dictated by the scalability of the blockchain---which is beyond the scope of this paper.
However, a recent study \cite{blackshear2023sui} shows that the blockchain we make use of supports thousands of transactions per second.

\subsection{Implementation}

\para{Control-Plane Smart Contracts}
We implemented our control plane as a set of smart contracts on top of the Sui blockchain~\cite{blackshear2023sui} that provide the bandwidth asset functionality, as well as a marketplace that allows buying and selling assets.

\para{Market Client Application}
To allow ASes and end hosts to interact with the smart contracts (i.e., the assets and marketplace), we implemented a client application written in Rust.
Our application allows an AS to create assets and marketplace listings and handles the assignment of $\resid$s (using an online First Fit algorithm~\cite{kierstead2016first,gyarfas1988line}), the derivation of the authentication keys, and the delivery of the reservation information (via smart contract call).


On the end host, the application handles buying and redeeming of assets and integrates with the \name data plane, i.e., delivers all necessary information to the applications using the reservation.

\para{Blockchain Platform \& Atomic Transactions}
We chose the Sui blockchain for our implementation due to its high throughput and low latency, which make it suitable for applications such as bandwidth reservations.
Sui transactions using only \emph{owned} objects benefit from a fast path (using Byzantine consistent broadcast). This reduces latency compared to accessing \emph{shared} objects, which require consensus.
In our design, all operations on the assets themselves can be performed using this fast path. Only interactions with a market require the consensus path.


\begin{table}[t]
	\centering
	\caption{Gas and dollar cost, rounded to two significant figures, of transactions to atomically buy and redeem a full path. The total cost in SUI is computed as \texttt{computation cost + storage cost - storage rebate}. 
	}\label{tab:gas_cost}
	\sisetup{exponent-mode = fixed, fixed-exponent = 0}
	\begin{tabular}{@{}r S[table-format=1.5] S[table-format=1.3] S[table-format=1.3] S[table-format=1.3] S[table-format=1.3]@{}}
		\toprule
		\textbf{Hops}& \textbf{Computation} & \multicolumn{2}{c}{\textbf{Storage (SUI)}} & \multicolumn{2}{c}{\textbf{Total}}\\
		\cmidrule(lr){3-4}  \cmidrule(lr){5-6}
		& \textbf{(SUI)} & \textbf{Cost} & \textbf{Rebate} & \textbf{SUI} & \textbf{USD} \\
		\midrule
		1 & 0.00075 & 0.047 & 0.016 & 0.031 & 0.038 \\
		2 & 0.00075 & 0.090 & 0.029 & 0.062 & 0.076 \\
		4 & 0.00075 & 0.18 & 0.054 & 0.12 & 0.15 \\
		8 & 0.0015 & 0.35 & 0.10 & 0.25 & 0.30 \\
		16 & 0.0030 & 0.69 & 0.20 & 0.49 & 0.60 \\
		\bottomrule
		\multicolumn{6}{l}{\footnotesize \sisetup{exponent-mode=input}Computation price: \num{7.5e-07} SUI/unit; storage price: \num{7.6e-06} SUI/byte;}\\ \multicolumn{6}{l}{\footnotesize \sisetup{exponent-mode=input}SUI price: \num{1.221} USD (as of 2024-04-18 14:09 UTC); }  \\
	\end{tabular}
\end{table}

\para{Evaluation}
Our benchmarks measure the cost incurred by the smart contract execution for atomically buying and redeeming reservations for an end-to-end path, as well as the end-to-end latency between initiating the buy-and-redeem transaction, until all reservations are delivered.
We perform the latency measurements on the Sui \texttt{testnet} which is globally replicated with a distribution that closely resembles its main blockchain (Sui \texttt{mainnet}).

\subsection{Results}

\para{Contract Execution Cost}
The cost for transactions in Sui is split into three components.
The \emph{computation cost} is charged according to fixed buckets of computational units, based on the computation complexity, which are then converted to a value in Sui according to the \emph{computation gas price} provided in the transaction.
In our results, the values are provided based on the current reference gas price on mainnet of \num{7.5e-07} SUI per unit.
The \emph{storage cost} of objects is charged based on a \emph{storage gas price}, which is currently  \num{7.6e-06} SUI per byte.
The third component is a \emph{storage rebate}, which the transaction sender receives when deleting an object, consisting of 99\% of the original price paid for its storage. 

\Cref{tab:gas_cost} shows the cost of atomically purchasing paths of different lengths.
Since the cost is deterministic, the table does not include any measurement uncertainties.
The cost is dominated by the storage cost resulting from splitting an asset and re-listing the pieces that are not bought.
In our benchmark, we perform a worst-case split for each asset on the path,
consisting of two splits in the time dimension and one in the bandwidth dimension.
%

The cost depends linearly on the path length, since a longer path involves more assets that are split off from assets listed in the market. The transaction fees are approximately \num{0.038} USD per hop, resulting in a total cost of \num{0.60} USD for a path with 16 hops. 
Compared to transaction fees on, e.g., Ethereum where the average transaction fees are \num{8.16} USD\footnote{30-day trailing average transaction fee on 2024-04-18, from \url{https://etherscan.io/chart/avg-txfee-usd}} this is extremely cheap, and compares favorably even to centralized payment providers such as Paypal or Square, which charge a percentage of the price plus a fee of \num{0.49} and \num{0.30} USD, respectively~\cite{creditcardfees}.
Importantly, most of the transaction fees are later refunded to the ASes during the redemption of the reservation.
Thus, the fees can be partially priced in to the asset listings on the marketplace.

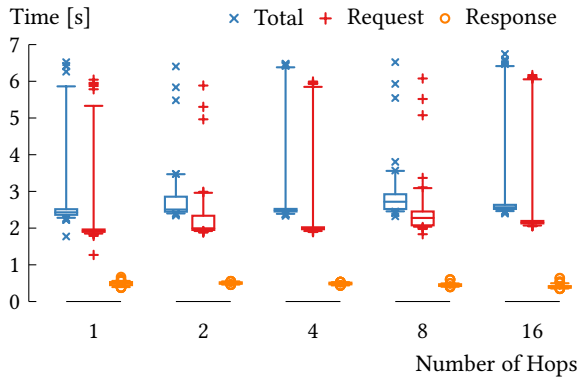
\begin{figure}[t]
  \centering
  \begin{tikzpicture}
  \begin{axis}[
    boxplot/draw direction=y,
    boxplot/every box/.style={solid},
    boxplot/every whisker/.style={solid},
    boxplot/every median/.style={solid},
    ytick={0,1,2,3,4,5,6,7},
    xtick={1 , 5, 9 ,13, 17},
    xticklabels={1, 2, 4, 8 , 16},
    xlabel={Number of Hops},
    ylabel={Time [s]},
    x tick style={draw=none},
    x axis line style={draw=none},
    ymin=0, ymax=7,
    cycle list name={shortlist},
    legend cell align=left,
    legend entries = {Total, Request, Response},
    legend pos = outer north east,
    legend style={at={(1,1.02)},anchor=south east, column sep=0.15cm},
    legend columns = 3,
    x tick label style={yshift={+5pt}},
    ]
\addlegendimage{only marks,mark=x, ctotal, mark options={fill=ctotal!50}}
\addlegendimage{only marks,mark=+, crequest, mark options={fill=crequest!50}}
\addlegendimage{only marks,mark=o, cresponse}

\draw (0,0) -- (2,0);
\draw (4,0) -- (6,0);
\draw (8,0) -- (10,0);
\draw (12,0) -- (14,0);
\draw (16,0) -- (18,0);

\addplot+ [
    crequest,
    mark=+,
    boxplot prepared={
    lower whisker=1.851,
    lower quartile=1.898,
    median=1.925,
    upper quartile=1.963,
    upper whisker=5.331,
    draw position=1,
},
] table [row sep=\\,y index=0] { 1.269065208\\ 1.845309167\\ 1.823399167\\ 1.787284042\\ 1.842883791\\ 5.871069542\\ 6.04382925\\ 5.956826041\\ 5.922371\\ 5.782925458\\ };

\addplot+ [
    cresponse,
    mark=o,
    boxplot prepared={
    lower whisker=0.394,
    lower quartile=0.447,
    median=0.500,
    upper quartile=0.530,
    upper whisker=0.561,
    draw position=2,
},
] table [row sep=\\,y index=0] { 0.374402708\\ 0.374444708\\ 0.378952\\ 0.37876975\\ 0.388884959\\ 0.563867166\\ 0.563325625\\ 0.56135975\\ 0.668366708\\ 0.611214541\\ };

\addplot+ [
    ctotal,
    mark=x,
    boxplot prepared={
    lower whisker=2.281,
    lower quartile=2.362,
    median=2.430,
    upper quartile=2.515,
    upper whisker=5.862,
    draw position=0,
},
] table [row sep=\\,y index=0] { 1.771932042\\ 2.225623542\\ 2.219351834\\ 2.256335458\\ 2.271133125\\ 6.40669475\\ 6.42278125\\ 6.520151666\\ 6.411704375\\ 6.261889916\\ };

\addplot+ [
    crequest,
    mark=+,
    boxplot prepared={
    lower whisker=1.921,
    lower quartile=1.950,
    median=1.994,
    upper quartile=2.337,
    upper whisker=2.962,
    draw position=5,
},
] table [row sep=\\,y index=0] { 1.909755334\\ 1.884695458\\ 1.919473709\\ 1.882315125\\ 1.916819125\\ 5.884117917\\ 2.975267125\\ 2.989503625\\ 5.303664417\\ 4.964084958\\ };

\addplot+ [
    cresponse,
    mark=o,
    boxplot prepared={
    lower whisker=0.462,
    lower quartile=0.483,
    median=0.502,
    upper quartile=0.524,
    upper whisker=0.548,
    draw position=6,
},
] table [row sep=\\,y index=0] { 0.456364291\\ 0.456388791\\ 0.459407709\\ 0.460161625\\ 0.461148708\\ 0.552887375\\ 0.555452333\\ 0.547948\\ 0.550744959\\ 0.553328416\\ };

\addplot+ [
    ctotal,
    mark=x,
    boxplot prepared={
    lower whisker=2.400,
    lower quartile=2.448,
    median=2.506,
    upper quartile=2.856,
    upper whisker=3.469,
    draw position=4,
},
] table [row sep=\\,y index=0] { 2.395627084\\ 2.389445625\\ 2.388283333\\ 2.341722834\\ 2.385581417\\ 6.403636708\\ 3.475962459\\ 3.476505333\\ 5.835961625\\ 5.48227375\\ };

\addplot+ [
    crequest,
    mark=+,
    boxplot prepared={
    lower whisker=1.921,
    lower quartile=1.970,
    median=1.994,
    upper quartile=2.022,
    upper whisker=5.848,
    draw position=9,
},
] table [row sep=\\,y index=0] { 1.919239459\\ 1.896419042\\ 1.8956035\\ 1.894747584\\ 1.894087541\\ 5.932392291\\ 5.949629334\\ 5.9612384590000005\\ 5.931827958\\ 5.993398833\\ };

\addplot+ [
    cresponse,
    mark=o,
    boxplot prepared={
    lower whisker=0.439,
    lower quartile=0.468,
    median=0.491,
    upper quartile=0.514,
    upper whisker=0.528,
    draw position=10,
},
] table [row sep=\\,y index=0] { 0.435121667\\ 0.438250958\\ 0.433427125\\ 0.436279542\\ 0.4383315\\ 0.533874417\\ 0.533469833\\ 0.533159333\\ 0.532085542\\ 0.532450959\\ };

\addplot+ [
    ctotal,
    mark=x,
    boxplot prepared={
    lower whisker=2.389,
    lower quartile=2.453,
    median=2.486,
    upper quartile=2.523,
    upper whisker=6.381,
    draw position=8,
},
] table [row sep=\\,y index=0] { 2.372317834\\ 2.378182459\\ 2.384106458\\ 2.33410625\\ 2.372939334\\ 6.446369833\\ 6.421124125\\ 6.46518125\\ 6.423227667\\ 6.480463125\\ };

\addplot+ [
    crequest,
    mark=+,
    boxplot prepared={
    lower whisker=2.038,
    lower quartile=2.076,
    median=2.277,
    upper quartile=2.453,
    upper whisker=3.090,
    draw position=13,
},
] table [row sep=\\,y index=0] { 1.976252958\\ 2.0088555\\ 1.833976125\\ 2.028585125\\ 2.003260167\\ 3.11079\\ 6.07624375\\ 3.367408125\\ 5.075298625\\ 5.517668458\\ };

\addplot+ [
    cresponse,
    mark=o,
    boxplot prepared={
    lower whisker=0.403,
    lower quartile=0.424,
    median=0.445,
    upper quartile=0.471,
    upper whisker=0.493,
    draw position=14,
},
] table [row sep=\\,y index=0] { 0.397688083\\ 0.401250084\\ 0.402362542\\ 0.401811541\\ 0.392699792\\ 0.493368917\\ 0.596752041\\ 0.498313041\\ 0.600150958\\ 0.602938667\\ };

\addplot+ [
    ctotal,
    mark=x,
    boxplot prepared={
    lower whisker=2.457,
    lower quartile=2.520,
    median=2.718,
    upper quartile=2.923,
    upper whisker=3.559,
    draw position=12,
},
] table [row sep=\\,y index=0] { 2.407265125\\ 2.418510167\\ 2.320600542\\ 2.452573708\\ 2.413580084\\ 6.521482042\\ 3.805211167\\ 5.544885792\\ 5.928710708\\ 3.568032583\\ };

\addplot+ [
    crequest,
    mark=+,
    boxplot prepared={
    lower whisker=2.071,
    lower quartile=2.128,
    median=2.156,
    upper quartile=2.195,
    upper whisker=6.054,
    draw position=17,
},
] table [row sep=\\,y index=0] { 2.069344\\ 2.060668375\\ 2.059691375\\ 2.049131625\\ 2.058589083\\ 6.116560041\\ 6.104704209\\ 6.1698485\\ 6.072207334\\ 6.076017083\\ };

\addplot+ [
    cresponse,
    mark=o,
    boxplot prepared={
    lower whisker=0.354,
    lower quartile=0.374,
    median=0.405,
    upper quartile=0.420,
    upper whisker=0.498,
    draw position=18,
},
] table [row sep=\\,y index=0] { 0.345394625\\ 0.351731208\\ 0.348704709\\ 0.346957208\\ 0.346136708\\ 0.53837825\\ 0.509063\\ 0.638612416\\ 0.632320458\\ 0.539437416\\ };

\addplot+ [
    ctotal,
    mark=x,
    boxplot prepared={
    lower whisker=2.464,
    lower quartile=2.522,
    median=2.566,
    upper quartile=2.634,
    upper whisker=6.414,
    draw position=16,
},
] table [row sep=\\,y index=0] { 2.414738625\\ 2.44646925\\ 2.46040475\\ 2.455353833\\ 2.396088833\\ 6.541302916\\ 6.743316625\\ 6.593011833\\ 6.556823125\\ 6.467503916\\ };

    \end{axis}
\end{tikzpicture}

  \caption{%
  End-to-end latency for an atomic buy-and-redeem (reporting total, request, and response latencies) for different path lengths, each measured 100 times.
 Time is measured from the point when the buyer has chosen which assets to buy until all of the reservation information required for the data plane has been delivered to the buyer.
    	The whiskers in the box plots represent the 5th and 95th percentiles.}
    \label{fig:e2e_timing}
\end{figure}

\para{End-to-end Latency}
\Cref{fig:e2e_timing} shows the end-to-end latency for atomically buying reservations for a full path. The measurements are run 100 times for each path length.
The request time describes the latency between initiating the purchase and the finalization of the purchase transaction.
The response time measures the latency from that point until the buyer has received the reservation information from all on-path ASes.
The responses make use of Sui's fast path, which lowers the latency significantly, whereas the purchase transaction interacts with a shared object (namely the marketplace), which requires going through consensus.
The latency is largely independent of the length of the path with a total latency of less than 3 seconds in 83\% of our measurements.

This end-to-end latency 
is comparable to the time required to authorize credit-card payments (``a few seconds''~\cite{stripelatency}).
Thus, any centralized bandwidth market that makes use of credit-card payments---independent of the underlying QoS system---is unlikely to provide a significantly lower latency.
For many applications, e.g., video calls, this should even be fast enough for obtaining reservations ad-hoc, e.g., when congestion occurs unexpectedly. However, \name also allows reservations to be established ahead of time---which we expect to be the common usage---in which case the latency matters less and is more than fast enough for any application.


\section{Data-Plane Evaluation}
\label{sec:evaluation_dataplane}

We implement two versions of the \name data plane.
The first extends the open-source SCION implementation in Golang, adding the \name data plane and providing an API for the control plane.
The second is a high-speed variant, which comprises the generation of reservation traffic at the source and the packet validation and forwarding at border routers. We use this latter version to demonstrate \name's potential for performance and compare it to a SCION-only data plane, constituting our baseline. We provide additional evaluation results in \cref{app:additional}.

\subsection{Implementation and Test Setup}
We implement both SCION and \name{} in DPDK~\cite{dpdk}, which is also used for SCION routers deployed in commercial networks~\cite{dpdk_scion_border_router}.
For the MAC and PRF operations, we use AES-128, taking advantage of Intel's AES-NI~\cite{AESni} hardware instructions. 
To deterministically police flyover reservations in \name, we use an array that establishes a direct mapping between the reservation ID and its corresponding 8B token bucket.
We initialize the array to store at most $10^5$ reservation IDs, requiring \SI{800}{kB} of memory.


Our testing environment consists of two machines: one off-the-shelf server with an Intel Xeon \SI{2.1}{GHz} CPU running Ubuntu and executing our implementation, and a Spirent SPT-N4U device.
The Spirent machine serves a dual role as a traffic generator during border router traffic validation evaluations, and as a bandwidth monitor when assessing traffic generation at a source.
Both machines are interconnected by four bidirectional Ethernet links operating at \SI{40}{Gbps} each.
We assess each of the two implementations independently.
When evaluating \name{}, we always measure its worst-case performance by assuming the existence of a flyover reservation at every on-path AS.

\subsection{Results}

\para{Traffic Forwarding}
The border router's validation and forwarding performance are shown in \cref{fig:dp_eval_router_tp_c}.
With only 4 cores, our implementation achieves the line rate of \SI{160}{Gbps} for \SI{1500}{\byte} payloads, and scales to send \SI{100}{\byte} payloads at line rate using \num{32} cores.
%
As anticipated, operations such as computing the SCION hop field MAC, authentication key, flyover MAC, and performing the overuse check emerge as the most resource-intensive operations.
We find that best-effort SCION packets can be processed in \SI{123}{ns}, while \name{} packets require \SI{308}{ns}.
Although this difference is noticeable, this does not imply a lower throughput in practice, as it can be mitigated by increasing the number of CPU cores.

\begin{figure}[t]
    \input{figures/dp_eval_router_tp_c}
    \caption{Border router packet validation and forwarding performance for different payload sizes and number of CPU cores.
    Solid lines correspond to \name reservations, dashed lines to standard SCION best-effort traffic. Note that the dashed lines for 1000B and 1500B are hidden behind the solid line for 1500B since the performance is virtually identical.}
    \label{fig:dp_eval_router_tp_c}
\end{figure}


\section{Conclusion}
\label{sec:conclusion}

We explore how the combination of an out-of-band control plane---in our case in the form of a smart contract running on a blockchain---with a data plane for inter-domain bandwidth reservations can result in a highly flexible, efficient, and fair system to enforce availability guarantees on a public Internet.

The efficient data plane makes \name easy to deploy on routers. The reservation model---whereby reservations are independently granted on each AS-level hop---facilitates incremental deployment, as not all on-path ASes need to offer bandwidth reservations. Further, the reservation market dynamics incentivize ASes and provide them with an additional source of revenue.

In conclusion, \name is the first practical inter-domain bandwidth-reservation system:
Its substantial benefits for endhosts, and clear economic incentives for ASes facilitate adoption, enabling applications with strong availability requirements to flourish on the SCION Internet.

\section*{Ethical Considerations} 

This paper does not raise any ethical concerns.

\section*{Acknowledgments}

We gratefully acknowledge support for parts of this project from the Werner Siemens Stiftung Centre for Cyber Trust at ETH Zurich.

\clearpage
\bibliographystyle{ACM-Reference-Format}
\bibliography{references}


\begin{thebibliography}{58}


\ifx \showCODEN    \undefined \def \showCODEN     #1{\unskip}     \fi
\ifx \showISBNx    \undefined \def \showISBNx     #1{\unskip}     \fi
\ifx \showISBNxiii \undefined \def \showISBNxiii  #1{\unskip}     \fi
\ifx \showISSN     \undefined \def \showISSN      #1{\unskip}     \fi
\ifx \showLCCN     \undefined \def \showLCCN      #1{\unskip}     \fi
\ifx \shownote     \undefined \def \shownote      #1{#1}          \fi
\ifx \showarticletitle \undefined \def \showarticletitle #1{#1}   \fi
\ifx \showURL      \undefined \def \showURL       {\relax}        \fi
\providecommand\bibfield[2]{#2}
\providecommand\bibinfo[2]{#2}
\providecommand\natexlab[1]{#1}
\providecommand\showeprint[2][]{arXiv:#2}

\bibitem[{Anapaya Systems}(2020)]%
        {dpdk_scion_border_router}
\bibfield{author}{\bibinfo{person}{{Anapaya Systems}}.}
  \bibinfo{year}{2020}\natexlab{}.
\newblock \bibinfo{title}{{SCION-Internet and Anapaya Software}}.
\newblock \bibinfo{howpublished}{\url{
  https://content.anapaya.net/hubfs/collateral/anapaya-scion-Internet-and-anapaya-software-fs-en.pdf?hsLang=en
  }}.
\newblock


\bibitem[Anderson et~al\mbox{.}(2004)]%
        {anderson2004preventing}
\bibfield{author}{\bibinfo{person}{Tom Anderson}, \bibinfo{person}{Timothy
  Roscoe}, {and} \bibinfo{person}{David Wetherall}.}
  \bibinfo{year}{2004}\natexlab{}.
\newblock \showarticletitle{Preventing {Internet} denial-of-service with
  capabilities}.
\newblock \bibinfo{journal}{\emph{ACM SIGCOMM Computer Communication Review
  (CCR)}} \bibinfo{volume}{34}, \bibinfo{number}{1} (\bibinfo{year}{2004}),
  \bibinfo{pages}{39–44}.
\newblock
\href{https://doi.org/10.1145/972374.972382}{doi:\nolinkurl{10.1145/972374.972382}}


\bibitem[Baker et~al\mbox{.}(1998)]%
        {rfc2474}
\bibfield{author}{\bibinfo{person}{Fred Baker}, \bibinfo{person}{David~L.
  Black}, \bibinfo{person}{Kathleen Nichols}, {and} \bibinfo{person}{Steven~L.
  Blake}.} \bibinfo{year}{1998}\natexlab{}.
\newblock \bibinfo{booktitle}{\emph{{Definition of the Differentiated Services
  Field (DS Field) in the IPv4 and IPv6 Headers}}}.
\newblock \bibinfo{type}{RFC} 2474. \bibinfo{institution}{IETF}.
\newblock
\href{https://doi.org/10.17487/RFC2474}{doi:\nolinkurl{10.17487/RFC2474}}


\bibitem[Ballani et~al\mbox{.}(2007)]%
        {ballani2007study}
\bibfield{author}{\bibinfo{person}{Hitesh Ballani}, \bibinfo{person}{Paul
  Francis}, {and} \bibinfo{person}{Xinyang Zhang}.}
  \bibinfo{year}{2007}\natexlab{}.
\newblock \showarticletitle{A study of prefix hijacking and interception in the
  {Internet}}.
\newblock \bibinfo{journal}{\emph{ACM SIGCOMM Computer Communication Review
  (CCR)}} \bibinfo{volume}{37}, \bibinfo{number}{4} (\bibinfo{year}{2007}),
  \bibinfo{pages}{265--276}.
\newblock
\href{https://doi.org/10.1145/1282380.1282411}{doi:\nolinkurl{10.1145/1282380.1282411}}


\bibitem[Basescu et~al\mbox{.}(2016)]%
        {basescu2016sibra}
\bibfield{author}{\bibinfo{person}{Cristina Basescu},
  \bibinfo{person}{Raphael~M. Reischuk}, \bibinfo{person}{Pawel Szalachowski},
  \bibinfo{person}{Adrian Perrig}, \bibinfo{person}{Yao Zhang},
  \bibinfo{person}{Hsu-Chun Hsiao}, \bibinfo{person}{Ayumu Kubota}, {and}
  \bibinfo{person}{Jumpei Urakawa}.} \bibinfo{year}{2016}\natexlab{}.
\newblock \showarticletitle{{SIBRA}: Scalable {Internet} Bandwidth Reservation
  Architecture}. In \bibinfo{booktitle}{\emph{Symposium on Network and
  Distributed Systems Security (NDSS)}}. \bibinfo{publisher}{The Internet
  Society}, \bibinfo{address}{Reston, VA, USA}, \bibinfo{numpages}{16}~pages.
\newblock
\href{https://doi.org/10.14722/ndss.2016.23132}{doi:\nolinkurl{10.14722/ndss.2016.23132}}


\bibitem[Blackshear et~al\mbox{.}(2023)]%
        {blackshear2023sui}
\bibfield{author}{\bibinfo{person}{Sam Blackshear}, \bibinfo{person}{Andrey
  Chursin}, \bibinfo{person}{George Danezis}, \bibinfo{person}{Anastasios
  Kichidis}, \bibinfo{person}{Lefteris Kokoris-Kogias}, \bibinfo{person}{Xun
  Li}, \bibinfo{person}{Mark Logan}, \bibinfo{person}{Ashok Menon},
  \bibinfo{person}{Todd Nowacki}, \bibinfo{person}{Alberto Sonnino},
  \bibinfo{person}{Brandon William}, {and} \bibinfo{person}{Lu Zhang}.}
  \bibinfo{year}{2023}\natexlab{}.
\newblock \bibinfo{booktitle}{\emph{{Sui Lutris}: A Blockchain Combining
  Broadcast and Consensus}}.
\newblock \bibinfo{type}{{T}echnical {R}eport}. \bibinfo{institution}{Mysten
  Labs}.
\newblock
\urldef\tempurl%
\url{https://arxiv.org/pdf/2310.18042}
\showURL{%
\tempurl}


\bibitem[Braden et~al\mbox{.}(1997)]%
        {rfc2205}
\bibfield{author}{\bibinfo{person}{Robert~T. Braden}, \bibinfo{person}{Lixia
  Zhang}, \bibinfo{person}{Steven Berson}, \bibinfo{person}{Shai Herzog}, {and}
  \bibinfo{person}{Sugih Jamin}.} \bibinfo{year}{1997}\natexlab{}.
\newblock \bibinfo{booktitle}{\emph{{Resource ReSerVation Protocol (RSVP) --
  Version 1 Functional Specification}}}.
\newblock \bibinfo{type}{RFC} 2205. \bibinfo{institution}{IETF}.
\newblock
\href{https://doi.org/10.17487/RFC2205}{doi:\nolinkurl{10.17487/RFC2205}}


\bibitem[Castro et~al\mbox{.}(2015)]%
        {castro2015route}
\bibfield{author}{\bibinfo{person}{Ignacio Castro}, \bibinfo{person}{Aurojit
  Panda}, \bibinfo{person}{Barath Raghavan}, \bibinfo{person}{Scott Shenker},
  {and} \bibinfo{person}{Sergey Gorinsky}.} \bibinfo{year}{2015}\natexlab{}.
\newblock \showarticletitle{Route bazaar: Automatic interdomain contract
  negotiation}. In \bibinfo{booktitle}{\emph{USENIX Workshop on Hot Topics in
  Operating Systems}}.
\newblock


\bibitem[Chuat et~al\mbox{.}(2022)]%
        {chuat2022complete}
\bibfield{author}{\bibinfo{person}{Laurent Chuat}, \bibinfo{person}{Markus
  Legner}, \bibinfo{person}{David Basin}, \bibinfo{person}{David Hausheer},
  \bibinfo{person}{Samuel Hitz}, \bibinfo{person}{Peter M{\"u}ller}, {and}
  \bibinfo{person}{Adrian Perrig}.} \bibinfo{year}{2022}\natexlab{}.
\newblock \bibinfo{booktitle}{\emph{The Complete Guide to {SCION}}}.
\newblock \bibinfo{publisher}{Springer}, \bibinfo{address}{Heidelberg, DE}.
\newblock
\href{https://doi.org/10.1007/978-3-031-05288-0}{doi:\nolinkurl{10.1007/978-3-031-05288-0}}


\bibitem[Cisco(2016)]%
        {ciscovoip}
\bibfield{author}{\bibinfo{person}{Cisco}.} \bibinfo{year}{2016}\natexlab{}.
\newblock \bibinfo{title}{Voice Packetization}.
\newblock
\urldef\tempurl%
\url{https://www.cisco.com/c/en/us/td/docs/ios-xml/ios/voice/cube/configuration/cube-book/cube-book_chapter_01001111.html}
\showURL{%
Retrieved August 16, 2023 from \tempurl}


\bibitem[Clarke(1971)]%
        {clarke1971multipart}
\bibfield{author}{\bibinfo{person}{Edward~H Clarke}.}
  \bibinfo{year}{1971}\natexlab{}.
\newblock \showarticletitle{Multipart pricing of public goods}.
\newblock \bibinfo{journal}{\emph{Public choice}} (\bibinfo{year}{1971}),
  \bibinfo{pages}{17--33}.
\newblock


\bibitem[Cocchi et~al\mbox{.}(1993)]%
        {cocchi1993pricing}
\bibfield{author}{\bibinfo{person}{R. Cocchi}, \bibinfo{person}{S. Shenker},
  \bibinfo{person}{D. Estrin}, {and} \bibinfo{person}{Lixia Zhang}.}
  \bibinfo{year}{1993}\natexlab{}.
\newblock \showarticletitle{Pricing in computer networks: motivation,
  formulation, and example}.
\newblock \bibinfo{journal}{\emph{IEEE/ACM Transactions on Networking}}
  \bibinfo{volume}{1}, \bibinfo{number}{6} (\bibinfo{year}{1993}),
  \bibinfo{pages}{614--627}.
\newblock
\href{https://doi.org/10.1109/90.266050}{doi:\nolinkurl{10.1109/90.266050}}


\bibitem[{DPDK Project}(2023)]%
        {dpdk}
\bibfield{author}{\bibinfo{person}{{DPDK Project}}.}
  \bibinfo{year}{2023}\natexlab{}.
\newblock \bibinfo{title}{{Data Plane Development Kit}}.
\newblock \bibinfo{howpublished}{\url{https://dpdk.org}}.
\newblock


\bibitem[et~al.(2025)]%
        {wirz2025sciera}
\bibfield{author}{\bibinfo{person}{Wirz et al.}}
  \bibinfo{year}{2025}\natexlab{}.
\newblock \showarticletitle{Scaling SCIERA: A Journey Through the Deployment of
  a Next-generation Network}. In \bibinfo{booktitle}{\emph{ACM SIGCOMM
  Conference}}. \bibinfo{publisher}{Association for Computing Machinery},
  \bibinfo{address}{New York, NY, USA}.
\newblock


\bibitem[Filsfils et~al\mbox{.}(2021)]%
        {SRv6}
\bibfield{author}{\bibinfo{person}{Clarence Filsfils}, \bibinfo{person}{Pablo
  Camarillo}, \bibinfo{person}{John Leddy}, \bibinfo{person}{Daniel Voyer},
  \bibinfo{person}{Satoru Matsushima}, {and} \bibinfo{person}{Zhenbin Li}.}
  \bibinfo{year}{2021}\natexlab{}.
\newblock \bibinfo{title}{{Segment Routing over IPv6 (SRv6) Network
  Programming}}.
\newblock \bibinfo{howpublished}{RFC 8986}.
\newblock
\href{https://doi.org/10.17487/RFC8986}{doi:\nolinkurl{10.17487/RFC8986}}


\bibitem[Fu and Modiano(2021)]%
        {Fu2021learning}
\bibfield{author}{\bibinfo{person}{Xinzhe Fu} {and} \bibinfo{person}{Eytan
  Modiano}.} \bibinfo{year}{2021}\natexlab{}.
\newblock \showarticletitle{Learning-NUM: Network Utility Maximization with
  Unknown Utility Functions and Queueing Delay}. In
  \bibinfo{booktitle}{\emph{Proceedings of the Twenty-Second International
  Symposium on Theory, Algorithmic Foundations, and Protocol Design for Mobile
  Networks and Mobile Computing}} (Shanghai, China)
  \emph{(\bibinfo{series}{MobiHoc '21})}. \bibinfo{publisher}{Association for
  Computing Machinery}, \bibinfo{address}{New York, NY, USA},
  \bibinfo{pages}{21–30}.
\newblock
\showISBNx{9781450385589}
\href{https://doi.org/10.1145/3466772.3467031}{doi:\nolinkurl{10.1145/3466772.3467031}}


\bibitem[Giuliari et~al\mbox{.}(2021a)]%
        {giuliari2021colibri}
\bibfield{author}{\bibinfo{person}{Giacomo Giuliari}, \bibinfo{person}{Dominik
  Roos}, \bibinfo{person}{Marc Wyss}, \bibinfo{person}{Juan~Angel Garc{\'\i}~a
  Pardo}, \bibinfo{person}{Markus Legner}, {and} \bibinfo{person}{Adrian
  Perrig}.} \bibinfo{year}{2021}\natexlab{a}.
\newblock \showarticletitle{Colibri: a cooperative lightweight inter-domain
  bandwidth-reservation infrastructure}. In
  \bibinfo{booktitle}{\emph{Conference on Emerging Networking Experiments and
  Technologies (CoNEXT)}}. \bibinfo{publisher}{Association for Computing
  Machinery}, \bibinfo{address}{New York, NY, USA}, \bibinfo{pages}{104--118}.
\newblock
\href{https://doi.org/10.1145/3485983.3494871}{doi:\nolinkurl{10.1145/3485983.3494871}}


\bibitem[Giuliari et~al\mbox{.}(2021b)]%
        {giuliari2021gma}
\bibfield{author}{\bibinfo{person}{Giacomo Giuliari}, \bibinfo{person}{Marc
  Wyss}, \bibinfo{person}{Markus Legner}, {and} \bibinfo{person}{Adrian
  Perrig}.} \bibinfo{year}{2021}\natexlab{b}.
\newblock \showarticletitle{{GMA}: A Pareto Optimal Distributed
  Resource-Allocation Algorithm}. In \bibinfo{booktitle}{\emph{Proceedings of
  the International Colloquium on Structural Information and Communication
  Complexity (SIROCCO)}}.
\newblock
\href{https://doi.org/10.1007/978-3-030-79527-6_14}{doi:\nolinkurl{10.1007/978-3-030-79527-6_14}}


\bibitem[Groves(1973)]%
        {groves1973incentives}
\bibfield{author}{\bibinfo{person}{Theodore Groves}.}
  \bibinfo{year}{1973}\natexlab{}.
\newblock \showarticletitle{Incentives in teams}.
\newblock \bibinfo{journal}{\emph{Econometrica: Journal of the Econometric
  Society}} (\bibinfo{year}{1973}), \bibinfo{pages}{617--631}.
\newblock


\bibitem[Gueron(2010)]%
        {AESni}
\bibfield{author}{\bibinfo{person}{Shay Gueron}.}
  \bibinfo{year}{2010}\natexlab{}.
\newblock \bibinfo{booktitle}{\emph{Intel {Advanced} {Encryption} {Standard}
  ({AES}) new instructions set}}.
\newblock \bibinfo{type}{{T}echnical {R}eport}. \bibinfo{institution}{Intel
  Corporation}.
\newblock
\urldef\tempurl%
\url{https://www.intel.com.bo/content/dam/doc/white-paper/advanced-encryption-standard-new-instructions-set-paper.pdf}
\showURL{%
\tempurl}


\bibitem[Gy{\'a}rf{\'a}s and Lehel(1988)]%
        {gyarfas1988line}
\bibfield{author}{\bibinfo{person}{Andr{\'a}s Gy{\'a}rf{\'a}s} {and}
  \bibinfo{person}{Jen{\"o} Lehel}.} \bibinfo{year}{1988}\natexlab{}.
\newblock \showarticletitle{On-line and first fit colorings of graphs}.
\newblock \bibinfo{journal}{\emph{Journal of Graph theory}}
  \bibinfo{volume}{12}, \bibinfo{number}{2} (\bibinfo{year}{1988}),
  \bibinfo{pages}{217--227}.
\newblock


\bibitem[Heilman et~al\mbox{.}(2015)]%
        {heilman2015eclipse}
\bibfield{author}{\bibinfo{person}{Ethan Heilman}, \bibinfo{person}{Alison
  Kendler}, \bibinfo{person}{Aviv Zohar}, {and} \bibinfo{person}{Sharon
  Goldberg}.} \bibinfo{year}{2015}\natexlab{}.
\newblock \showarticletitle{Eclipse attacks on {Bitcoin}'s peer-to-peer
  network}. In \bibinfo{booktitle}{\emph{USENIX Security Symposium (USENIX
  Security)}}. \bibinfo{publisher}{USENIX Association},
  \bibinfo{address}{Berkeley, CA, USA}, \bibinfo{pages}{129--144}.
\newblock
\urldef\tempurl%
\url{https://www.usenix.org/conference/usenixsecurity15/technical-sessions/presentation/heilman}
\showURL{%
\tempurl}


\bibitem[Hsiao et~al\mbox{.}(2013)]%
        {Hsiao2013}
\bibfield{author}{\bibinfo{person}{Hsu-Chun Hsiao}, \bibinfo{person}{Tiffany
  Hyun-Jin Kim}, \bibinfo{person}{Sangjae Yoo}, \bibinfo{person}{Xin Zhang},
  \bibinfo{person}{Soo~Bum Lee}, \bibinfo{person}{Virgil Gligor}, {and}
  \bibinfo{person}{Adrian Perrig}.} \bibinfo{year}{2013}\natexlab{}.
\newblock \showarticletitle{{STRIDE}: Sanctuary Trail -- Refuge from {Internet}
  {DDoS} Entrapment.}. In \bibinfo{booktitle}{\emph{ACM Asia Conference on
  Computer and Communications Security (ASIACCS)}}.
\newblock
\href{https://doi.org/10.1145/2484313.2484367}{doi:\nolinkurl{10.1145/2484313.2484367}}


\bibitem[Jacobson(1988)]%
        {jacobson1988congestion}
\bibfield{author}{\bibinfo{person}{V. Jacobson}.}
  \bibinfo{year}{1988}\natexlab{}.
\newblock \showarticletitle{Congestion avoidance and control}.
\newblock \bibinfo{journal}{\emph{SIGCOMM Comput. Commun. Rev.}}
  \bibinfo{volume}{18}, \bibinfo{number}{4} (\bibinfo{date}{Aug.}
  \bibinfo{year}{1988}), \bibinfo{pages}{314–329}.
\newblock
\showISSN{0146-4833}
\href{https://doi.org/10.1145/52325.52356}{doi:\nolinkurl{10.1145/52325.52356}}


\bibitem[Kashyop et~al\mbox{.}(2020)]%
        {kashyop2020dynamic}
\bibfield{author}{\bibinfo{person}{Manas~Jyoti Kashyop}, \bibinfo{person}{NS
  Narayanaswamy}, {et~al\mbox{.}}} \bibinfo{year}{2020}\natexlab{}.
\newblock \showarticletitle{Dynamic data structures for interval coloring}.
\newblock \bibinfo{journal}{\emph{Theoretical Computer Science}}
  \bibinfo{volume}{838} (\bibinfo{year}{2020}), \bibinfo{pages}{126--142}.
\newblock


\bibitem[Katz and Lindell(2008)]%
        {katz2008aggregate}
\bibfield{author}{\bibinfo{person}{Jonathan Katz} {and}
  \bibinfo{person}{Andrew~Y Lindell}.} \bibinfo{year}{2008}\natexlab{}.
\newblock \showarticletitle{Aggregate message authentication codes}. In
  \bibinfo{booktitle}{\emph{Cryptographers’ Track at the RSA Conference}}.
  \bibinfo{publisher}{Springer}, \bibinfo{address}{Berlin, Heidelberg},
  \bibinfo{pages}{155--169}.
\newblock


\bibitem[Kelly et~al\mbox{.}(1998)]%
        {Kelly1998rate}
\bibfield{author}{\bibinfo{person}{F~P Kelly}, \bibinfo{person}{A~K Maulloo},
  {and} \bibinfo{person}{D~K~H Tan}.} \bibinfo{year}{1998}\natexlab{}.
\newblock \showarticletitle{Rate control for communication networks: shadow
  prices, proportional fairness and stability}.
\newblock \bibinfo{journal}{\emph{Journal of the Operational Research Society}}
  \bibinfo{volume}{49}, \bibinfo{number}{3} (\bibinfo{year}{1998}),
  \bibinfo{pages}{237--252}.
\newblock
\showeprint{https://doi.org/10.1057/palgrave.jors.2600523}
\href{https://doi.org/10.1057/palgrave.jors.2600523}{doi:\nolinkurl{10.1057/palgrave.jors.2600523}}


\bibitem[Kierstead et~al\mbox{.}(2016)]%
        {kierstead2016first}
\bibfield{author}{\bibinfo{person}{Hal~A Kierstead}, \bibinfo{person}{David~A
  Smith}, {and} \bibinfo{person}{William~T Trotter}.}
  \bibinfo{year}{2016}\natexlab{}.
\newblock \showarticletitle{First-fit coloring on interval graphs has
  performance ratio at least 5}.
\newblock \bibinfo{journal}{\emph{European Journal of Combinatorics}}
  \bibinfo{volume}{51} (\bibinfo{year}{2016}), \bibinfo{pages}{236--254}.
\newblock


\bibitem[Kierstead et~al\mbox{.}(1981)]%
        {kierstead1981extremal}
\bibfield{author}{\bibinfo{person}{Henry~A Kierstead},
  \bibinfo{person}{William~T Trotter}, {et~al\mbox{.}}}
  \bibinfo{year}{1981}\natexlab{}.
\newblock \showarticletitle{An extremal problem in recursive combinatorics}.
\newblock \bibinfo{journal}{\emph{Congressus Numerantium}}
  \bibinfo{volume}{33}, \bibinfo{number}{143-153} (\bibinfo{year}{1981}),
  \bibinfo{pages}{98}.
\newblock


\bibitem[Kim et~al\mbox{.}(2014)]%
        {kim2014lightweight}
\bibfield{author}{\bibinfo{person}{Tiffany Hyun-Jin Kim},
  \bibinfo{person}{Cristina Basescu}, \bibinfo{person}{Limin Jia},
  \bibinfo{person}{Soo~Bum Lee}, \bibinfo{person}{Yih-Chun Hu}, {and}
  \bibinfo{person}{Adrian Perrig}.} \bibinfo{year}{2014}\natexlab{}.
\newblock \showarticletitle{Lightweight source authentication and path
  validation}. In \bibinfo{booktitle}{\emph{ACM SIGCOMM Conference}}.
  \bibinfo{publisher}{Association for Computing Machinery},
  \bibinfo{address}{New York, NY, USA}, \bibinfo{pages}{271--282}.
\newblock
\href{https://doi.org/10.1145/2619239.2626323}{doi:\nolinkurl{10.1145/2619239.2626323}}


\bibitem[Lee and Gligor(2010)]%
        {Lee2010Floc}
\bibfield{author}{\bibinfo{person}{Soo~Bum Lee} {and}
  \bibinfo{person}{Virgil~D. Gligor}.} \bibinfo{year}{2010}\natexlab{}.
\newblock \showarticletitle{{FLoc}: Dependable link access for legitimate
  traffic in flooding attacks}. In \bibinfo{booktitle}{\emph{Proc. of the IEEE
  International Conference on Distributed Computing Systems (ICDCS)}}. IEEE.
\newblock


\bibitem[Lee et~al\mbox{.}(2013)]%
        {Lee2013CoDef}
\bibfield{author}{\bibinfo{person}{Soo~Bum Lee}, \bibinfo{person}{Min~Suk
  Kang}, {and} \bibinfo{person}{Virgil~D. Gligor}.}
  \bibinfo{year}{2013}\natexlab{}.
\newblock \showarticletitle{{CoDef}: Collaborative defense against large-scale
  link-flooding attacks}. In \bibinfo{booktitle}{\emph{Conference on Emerging
  Networking Experiments and Technologies (CoNEXT)}}.
\newblock


\bibitem[Lee et~al\mbox{.}(2017)]%
        {lee2017case}
\bibfield{author}{\bibinfo{person}{Taeho Lee}, \bibinfo{person}{Christos
  Pappas}, \bibinfo{person}{Adrian Perrig}, \bibinfo{person}{Virgil Gligor},
  {and} \bibinfo{person}{Yih-Chun Hu}.} \bibinfo{year}{2017}\natexlab{}.
\newblock \showarticletitle{The Case for In-Network Replay Suppression}. In
  \bibinfo{booktitle}{\emph{ACM Asia Conference on Computer and Communications
  Security (ASIACCS)}}. \bibinfo{publisher}{Association for Computing
  Machinery}, \bibinfo{address}{New York, NY, USA},
  \bibinfo{numpages}{12}~pages.
\newblock
\href{https://doi.org/10.1145/3052973.3052988}{doi:\nolinkurl{10.1145/3052973.3052988}}


\bibitem[Leonard and Cassie(2023)]%
        {creditcardfees}
\bibfield{author}{\bibinfo{person}{Kimberlee Leonard} {and}
  \bibinfo{person}{Bottorff Cassie}.} \bibinfo{year}{2023}\natexlab{}.
\newblock \bibinfo{title}{Credit Card Processing Fees (2024 Guide)}.
\newblock \bibinfo{howpublished}{Forbes Advisor}.
\newblock
\urldef\tempurl%
\url{https://www.forbes.com/advisor/business/credit-card-processing-fees/}
\showURL{%
Retrieved January 16, 2024 from \tempurl}


\bibitem[Lepinski and Kent(2012)]%
        {rfc6480}
\bibfield{author}{\bibinfo{person}{Matt Lepinski} {and}
  \bibinfo{person}{Stephen Kent}.} \bibinfo{year}{2012}\natexlab{}.
\newblock \bibinfo{booktitle}{\emph{{An Infrastructure to Support Secure
  Internet Routing}}}.
\newblock \bibinfo{type}{RFC} 6480. \bibinfo{institution}{IETF}.
\newblock
\href{https://doi.org/10.17487/RFC6480}{doi:\nolinkurl{10.17487/RFC6480}}


\bibitem[MacKie-Mason and Varian(1995)]%
        {mason1995pricingtheinternet}
\bibfield{author}{\bibinfo{person}{Jeffrey MacKie-Mason} {and}
  \bibinfo{person}{Hal Varian}.} \bibinfo{year}{1995}\natexlab{}.
\newblock \bibinfo{title}{Pricing the Internet}.
\newblock \bibinfo{howpublished}{{MIT} Press}.
\newblock


\bibitem[Neely(2013)]%
        {neely2013delay}
\bibfield{author}{\bibinfo{person}{Michael~J. Neely}.}
  \bibinfo{year}{2013}\natexlab{}.
\newblock \showarticletitle{Delay-Based Network Utility Maximization}.
\newblock \bibinfo{journal}{\emph{IEEE/ACM Transactions on Networking}}
  \bibinfo{volume}{21}, \bibinfo{number}{1} (\bibinfo{year}{2013}),
  \bibinfo{pages}{41--54}.
\newblock
\href{https://doi.org/10.1109/TNET.2012.2191157}{doi:\nolinkurl{10.1109/TNET.2012.2191157}}


\bibitem[Parno et~al\mbox{.}(2007)]%
        {Parno2007}
\bibfield{author}{\bibinfo{person}{Bryan Parno}, \bibinfo{person}{Dan
  Wendlandt}, \bibinfo{person}{Elaine Shi}, \bibinfo{person}{Adrian Perrig},
  \bibinfo{person}{Bruce Maggs}, {and} \bibinfo{person}{Yih-Chun Hu}.}
  \bibinfo{year}{2007}\natexlab{}.
\newblock \showarticletitle{Portcullis: Protecting Connection Setup from
  Denial-of-Capability Attacks}. In \bibinfo{booktitle}{\emph{ACM SIGCOMM
  Conference}}.
\newblock
\href{https://doi.org/10.1145/1282380.1282413}{doi:\nolinkurl{10.1145/1282380.1282413}}


\bibitem[PeeringDB(2023)]%
        {peeringdb}
\bibfield{author}{\bibinfo{person}{PeeringDB}.}
  \bibinfo{year}{2023}\natexlab{}.
\newblock \bibinfo{title}{{PeeringDB}}.
\newblock
\urldef\tempurl%
\url{https://www.peeringdb.com/}
\showURL{%
Retrieved August 16, 2023 from \tempurl}


\bibitem[P.T. and Sundaresan(2019)]%
        {akhil2019network}
\bibfield{author}{\bibinfo{person}{Akhil P.T.} {and} \bibinfo{person}{Rajesh
  Sundaresan}.} \bibinfo{year}{2019}\natexlab{}.
\newblock \showarticletitle{Network utility maximization revisited: Three
  issues and their resolution}.
\newblock \bibinfo{journal}{\emph{Performance Evaluation}}
  \bibinfo{volume}{136} (\bibinfo{year}{2019}), \bibinfo{pages}{102050}.
\newblock
\showISSN{0166-5316}
\href{https://doi.org/10.1016/j.peva.2019.102050}{doi:\nolinkurl{10.1016/j.peva.2019.102050}}


\bibitem[Reports(2023)]%
        {verifiedmarketreports-leased-line}
\bibfield{author}{\bibinfo{person}{Verified~Market Reports}.}
  \bibinfo{year}{2023}\natexlab{}.
\newblock \bibinfo{title}{Global Managed Leased Line Service Market By Type
  (Analog Dedicated Line, Digital Line), By Application (BFSI, Medical
  Insurance), By Geographic Scope And Forecast}.
\newblock
\urldef\tempurl%
\url{https://www.verifiedmarketreports.com/product/managed-leased-line-service-market/}
\showURL{%
Retrieved 2 September 2024 from \tempurl}


\bibitem[Research(2024)]%
        {databridge-sdwan}
\bibfield{author}{\bibinfo{person}{Databridge~Market Research}.}
  \bibinfo{year}{2024}\natexlab{}.
\newblock \bibinfo{title}{Global Software-Defined Wide Area Network ({SD-WAN})
  Market – Industry Trends and Forecast to 2031}.
\newblock
\urldef\tempurl%
\url{https://www.databridgemarketresearch.com/reports/global-software-defined-wide-area-network-sd-wan-market}
\showURL{%
Retrieved 2 September 2024 from \tempurl}


\bibitem[Rosen and Rekhter(2001)]%
        {rfc3107}
\bibfield{author}{\bibinfo{person}{Eric~C. Rosen} {and} \bibinfo{person}{Yakov
  Rekhter}.} \bibinfo{year}{2001}\natexlab{}.
\newblock \bibinfo{title}{{Carrying Label Information in BGP-4}}.
\newblock \bibinfo{howpublished}{RFC 3107}.
\newblock
\href{https://doi.org/10.17487/RFC3107}{doi:\nolinkurl{10.17487/RFC3107}}


\bibitem[Rothenberger et~al\mbox{.}(2020)]%
        {rothenberger2020piskes}
\bibfield{author}{\bibinfo{person}{Benjamin Rothenberger},
  \bibinfo{person}{Dominik Roos}, \bibinfo{person}{Markus Legner}, {and}
  \bibinfo{person}{Adrian Perrig}.} \bibinfo{year}{2020}\natexlab{}.
\newblock \showarticletitle{{PISKES}: Pragmatic {Internet}-scale
  key-establishment system}. In \bibinfo{booktitle}{\emph{ACM Asia Conference
  on Computer and Communications Security (ASIACCS)}}.
  \bibinfo{publisher}{Association for Computing Machinery},
  \bibinfo{address}{New York, NY, USA}, \bibinfo{pages}{73--86}.
\newblock
\href{https://doi.org/10.1145/3320269.3384743}{doi:\nolinkurl{10.1145/3320269.3384743}}


\bibitem[Stripe(2023)]%
        {stripelatency}
\bibfield{author}{\bibinfo{person}{Stripe}.} \bibinfo{year}{2023}\natexlab{}.
\newblock \bibinfo{title}{Card authorisation explained: How it works and what
  businesses need to know}.
\newblock
\urldef\tempurl%
\url{https://stripe.com/en-gb-ch/resources/more/card-authorization-explained}
\showURL{%
Retrieved January 17, 2024 from \tempurl}


\bibitem[Support(2023)]%
        {zoomrequirements}
\bibfield{author}{\bibinfo{person}{Zoom Support}.}
  \bibinfo{year}{2023}\natexlab{}.
\newblock \bibinfo{title}{Zoom system requirements: Windows, MacOS, Linux}.
\newblock
\urldef\tempurl%
\url{https://support.zoom.us/hc/en-us/articles/201362023-Zoom-system-requirements-Windows-macOS-Linux}
\showURL{%
Retrieved August 16, 2023 from \tempurl}


\bibitem[Vickrey(1961)]%
        {vickrey1961counterspeculation}
\bibfield{author}{\bibinfo{person}{William Vickrey}.}
  \bibinfo{year}{1961}\natexlab{}.
\newblock \showarticletitle{Counterspeculation, auctions, and competitive
  sealed tenders}.
\newblock \bibinfo{journal}{\emph{The Journal of finance}}
  \bibinfo{volume}{16}, \bibinfo{number}{1} (\bibinfo{year}{1961}),
  \bibinfo{pages}{8--37}.
\newblock


\bibitem[Viswanathan et~al\mbox{.}(2001)]%
        {mpls}
\bibfield{author}{\bibinfo{person}{Arun Viswanathan}, \bibinfo{person}{Eric~C.
  Rosen}, {and} \bibinfo{person}{Ross Callon}.}
  \bibinfo{year}{2001}\natexlab{}.
\newblock \bibinfo{title}{{Multiprotocol Label Switching Architecture}}.
\newblock \bibinfo{howpublished}{RFC 3031}.
\newblock
\href{https://doi.org/10.17487/RFC3031}{doi:\nolinkurl{10.17487/RFC3031}}


\bibitem[Werner et~al\mbox{.}(2022)]%
        {werner2021sok}
\bibfield{author}{\bibinfo{person}{Sam~M. Werner}, \bibinfo{person}{Daniel
  Perez}, \bibinfo{person}{Lewis Gudgeon}, \bibinfo{person}{Ariah
  Klages-Mundt}, \bibinfo{person}{Dominik Harz}, {and}
  \bibinfo{person}{William~J. Knottenbelt}.} \bibinfo{year}{2022}\natexlab{}.
\newblock \bibinfo{title}{{SoK}: Decentralized Finance ({DeFi})}.
\newblock
\showeprint[arxiv]{2101.08778}~[cs.CR]


\bibitem[Wikipedia(2024)]%
        {NTP-wikipedia}
\bibfield{author}{\bibinfo{person}{Wikipedia}.}
  \bibinfo{year}{2024}\natexlab{}.
\newblock \bibinfo{title}{Network Time Protocol}.
\newblock
\urldef\tempurl%
\url{https://en.wikipedia.org/wiki/Network_Time_Protocol}
\showURL{%
Retrieved 2 September 2024 from \tempurl}


\bibitem[Wroclawski(1997)]%
        {rfc2210}
\bibfield{author}{\bibinfo{person}{John~T. Wroclawski}.}
  \bibinfo{year}{1997}\natexlab{}.
\newblock \bibinfo{booktitle}{\emph{{The Use of RSVP with IETF Integrated
  Services}}}.
\newblock \bibinfo{type}{RFC} 2210. \bibinfo{institution}{IETF}.
\newblock
\href{https://doi.org/10.17487/RFC2210}{doi:\nolinkurl{10.17487/RFC2210}}


\bibitem[W{\"u}st and Gervais(2016)]%
        {wuest2016ethereum}
\bibfield{author}{\bibinfo{person}{Karl W{\"u}st} {and} \bibinfo{person}{Arthur
  Gervais}.} \bibinfo{year}{2016}\natexlab{}.
\newblock \bibinfo{booktitle}{\emph{Ethereum eclipse attacks}}.
\newblock \bibinfo{type}{{T}echnical {R}eport}. \bibinfo{institution}{ETH
  Zurich}.
\newblock
\href{https://doi.org/10.3929/ethz-a-010724205}{doi:\nolinkurl{10.3929/ethz-a-010724205}}


\bibitem[Wyss et~al\mbox{.}(2021)]%
        {wyss2021secure}
\bibfield{author}{\bibinfo{person}{Marc Wyss}, \bibinfo{person}{Giacomo
  Giuliari}, \bibinfo{person}{Markus Legner}, {and} \bibinfo{person}{Adrian
  Perrig}.} \bibinfo{year}{2021}\natexlab{}.
\newblock \showarticletitle{Secure and Scalable {QoS} for Critical
  Applications}. In \bibinfo{booktitle}{\emph{IEEE/ACM International Symposium
  on Quality of Service (IWQoS)}}.
\newblock


\bibitem[Wyss et~al\mbox{.}(2022)]%
        {wyss2022protecting}
\bibfield{author}{\bibinfo{person}{Marc Wyss}, \bibinfo{person}{Giacomo
  Giuliari}, \bibinfo{person}{Jonas Mohler}, {and} \bibinfo{person}{Adrian
  Perrig}.} \bibinfo{year}{2022}\natexlab{}.
\newblock \showarticletitle{Protecting Critical Inter-Domain Communication
  through Flyover Reservations}. In \bibinfo{booktitle}{\emph{ACM Conference on
  Computer and Communications Security (CCS)}}. \bibinfo{publisher}{Association
  for Computing Machinery}, \bibinfo{address}{New York, NY, USA},
  \bibinfo{pages}{2961--2974}.
\newblock
\href{https://doi.org/10.1145/3548606.3560582}{doi:\nolinkurl{10.1145/3548606.3560582}}


\bibitem[Wüst and Gervais(2018)]%
        {wuest2018blockchain}
\bibfield{author}{\bibinfo{person}{Karl Wüst} {and} \bibinfo{person}{Arthur
  Gervais}.} \bibinfo{year}{2018}\natexlab{}.
\newblock \showarticletitle{Do you Need a Blockchain?}. In
  \bibinfo{booktitle}{\emph{2018 Crypto Valley Conference on Blockchain
  Technology (CVCBT)}}.
\newblock
\href{https://doi.org/10.1109/CVCBT.2018.00011}{doi:\nolinkurl{10.1109/CVCBT.2018.00011}}


\bibitem[Yaar et~al\mbox{.}(2004)]%
        {Yaar2004SIFF}
\bibfield{author}{\bibinfo{person}{Abraham Yaar}, \bibinfo{person}{Adrian
  Perrig}, {and} \bibinfo{person}{Dawn Song}.} \bibinfo{year}{2004}\natexlab{}.
\newblock \showarticletitle{{SIFF}: A stateless {Internet} flow filter to
  mitigate {DDoS} flooding attacks}. In \bibinfo{booktitle}{\emph{IEEE
  Symposium on Security and Privacy (S\&P)}}.
\newblock


\bibitem[Yang et~al\mbox{.}(2005)]%
        {Yang2005}
\bibfield{author}{\bibinfo{person}{Xiaowei Yang}, \bibinfo{person}{David
  Wetherall}, {and} \bibinfo{person}{Thomas Anderson}.}
  \bibinfo{year}{2005}\natexlab{}.
\newblock \showarticletitle{A {DoS}-limiting network architecture}. In
  \bibinfo{booktitle}{\emph{ACM SIGCOMM Conference}}.
\newblock
\href{https://doi.org/10.1145/1080091.1080120}{doi:\nolinkurl{10.1145/1080091.1080120}}


\bibitem[Yeluri(2023)]%
        {sharada2023sizing}
\bibfield{author}{\bibinfo{person}{Sharada Yeluri}.}
  \bibinfo{year}{2023}\natexlab{}.
\newblock \bibinfo{title}{Sizing router buffers — small is the new big}.
\newblock
\urldef\tempurl%
\url{https://blog.apnic.net/2023/03/06/sizing-router-buffers-small-is-the-new-big/}
\showURL{%
Retrieved August 16, 2023 from \tempurl}


\end{thebibliography}

\appendix
\section*{Appendices}
Appendices are supporting material that has not been peer-reviewed.

\FloatBarrier

\section{Specification for \name on SCION}\label{sec:headerspec}

\newlength{\maxheight}
\setlength{\maxheight}{\heightof{W}}
\newcommand{\baselinealign}[1]{%
    \centering
    \raisebox{0pt}[\maxheight][0pt]{#1}%
}

In this appendix, we include the specification of a new SCION~\cite{chuat2022complete} path type that supports \name reservations.
SCION is well suited to deploy a bandwidth-reservation system such as ours, since it provides path choice to the source, which guarantees the stability of the path.
In contrast, path stability in the current Internet with BGP routing is only provided as long as there is convergence in the network.
For example, path stability is not guaranteed in the presence of BGP hijacking attacks~\cite{ballani2007study}.

This header specification is written analogously to the SCION header specification\footnote{\url{https://docs.scion.org/en/latest/protocols/scion-header.html}} and duplicates/reuses some of the information.
Changes compared to the SCION header specification are indicated.
The header layout for this path type is shown in \cref{fig:hdr:path}; the individual parts are described in further detail in the following subsections.

\begin{figure}
    \centering
    \begin{bytefield}[boxformatting=\baselinealign,bitwidth=0.03\columnwidth]{32}
        \wordbox{1}{PathMetaHdr}\\
        \wordbox{1}{InfoField}\\
        \wordbox{1}{\dots}\\
        \wordbox{1}{InfoField}\\
        \wordbox{1}{HopField}\\
        \wordbox{1}{HopField / FlyoverField}\\
        \wordbox{1}{\dots}
    \end{bytefield}
    \caption{Path header layout for the \name path type consisting of a path meta header, up to 3 info fields, and up to 64 hop fields or flyover fields.}
    \label{fig:hdr:path}
\end{figure}

\subsection{Path Meta Header (changed)}

\begin{figure}
    \centering
    \begin{bytefield}[boxformatting=\baselinealign,bitwidth=0.03\columnwidth]{32}
        \bitheader{0-31} \\
        \bitbox{2}{C} & \bitbox{8}{CurrHF} & \bitbox{1}{r} & \bitbox{7}{Seg0Len} & \bitbox{7}{Seg1Len} & \bitbox{7}{Seg2Len}\\
        \wordbox{1}{BaseTimestamp}\\
        \bitbox{10}{MillisTimestamp} & \bitbox{22}{Counter}
    \end{bytefield}
    \caption{Layout of the PathMetaHdr.}
    \label{fig:hdr:pathmetahdr}
\end{figure}

The PathMetaHdr is a 12-byte header containing meta information about the SCION path contained in the path header, see \cref{fig:hdr:pathmetahdr}.
It contains the following fields:
\begin{description}
    \item[(C)urrINF]
        2-bit index (0-based) pointing to the current info field (see offset calculations below).
    \item[CurrHF (changed)]
        8-bit index (0-based) pointing to the start of the current hop field (see offset calculations below) in 4-byte increments.
        This index is increased by 3 for normal hop fields and by 5 for flyover hop fields, which are \SI{12}{\byte} and \SI{20}{\byte} long, respectively.
    \item[r] Unused and reserved for future use.
    \item[Seg\{0,1,2\}Len (changed)]
        7-bit encoding of the length of each segment. The value in these fields is the length of the respective segment in bytes divided by 4.
        $Seg_iLen > 0$ implies the existence of info field $i$.
    \item[BaseTimestamp (new)]
        A unix timestamp (unsigned integer, 1-second granularity, similar to beacon timestamp in normal SCION path segments) that is used as a base to calculate start times for flyovers and the high granularity MillisTimestamp.
    \item[MillisTimestamp (new)]
        Millisecond granularity timestamp, as offset from BaseTimestamp. Used to compute MACs for flyover hops and to check recentness of a packet.
    \item[Counter (new)]
        A counter for each packet that is sent by the source to ensure that the tuple (BaseTimestamp, MillisTimestamp, Counter) is unique. This can then be used for the optional duplicate suppression at an AS.
\end{description}

\para{Path Offset Calculations (changed)}
The number of info fields is implied by $Seg_iLen > 0, i \in [0,2]$, thus $\fl{NumINF} = N + 1$ where $N = \max_{i \in [0,2]}$ s.t.\ $\forall j\leq i: \fl{Seg_jLen} > 0$.
It is an error to have $\fl{Seg_XLen} > 0 \wedge \fl{Seg_YLen} = 0$ for $X > Y$.
If all $\fl{Seg_iLen} = 0, i \in [0,2]$, then this denotes an empty path, which is only valid for intra-AS communication.

The offsets of the current info field and current hop field (relative to the end of the address header) are now calculated as
\begin{subequations}%
\begin{align}%
\fl{InfoFieldOffset} &= \SI{12}{\byte} + \SI{8}{\byte} \cdot \fl{CurrINF},\\
\fl{HopFieldOffset}  &= \SI{12}{\byte} + \SI{8}{\byte} \cdot \fl{NumINF} + \SI{4}{\byte} \cdot \fl{CurrHF}.
\end{align}
\end{subequations}

To check that the current hop field is in the segment of the current info field, the CurrHF needs to be compared to the SegLen fields of the current and preceding info fields.

\subsection{Info Field (unchanged)}

\begin{figure}
    \centering
    \begin{bytefield}[boxformatting=\baselinealign,bitwidth=0.03\columnwidth]{32}
        \bitheader{0-31} \\
        \bitboxes*{1}{rrrrrrPC}& \bitbox{8}{RSV} & \bitbox{16}{SegID} \\
        \wordbox{1}{Timestamp}
    \end{bytefield}
    \caption{Layout of an InfoField.}
    \label{fig:hdr:info}
\end{figure}

The format of an InfoField is the same as in the SCION path type and shown in \cref{fig:hdr:info}.
It contains the following fields:

\begin{description}
    \item[r] Unused and reserved for future use.
    \item[P]
        Peering flag.
        If set to true, then the forwarding path is built as a peering path, which requires special processing on the data plane.
    \item[C]
        Construction direction flag.
        If set to true then the hop fields are arranged in the direction they have been constructed during beaconing.
    \item[RSV]
        Unused and reserved for future use.
    \item[SegID]
        Updatable field used in the MAC-chaining mechanism.
    \item[Timestamp]
        Timestamp created by the initiator of the corresponding beacon.
        The timestamp is expressed in Unix time, and is encoded as an unsigned integer within 4~bytes with 1-second time granularity.
        It enables validation of the hop field by verification of the expiration time and MAC.
\end{description}

\subsection{HopField (slightly changed)}

\begin{figure}
    \centering
    \begin{bytefield}[boxformatting=\baselinealign,bitwidth=0.03\columnwidth]{32}
        \bitheader{0-31} \\
        \bitboxes*{1}{FrrrrrIE}& \bitbox{8}{ExpTime} & \bitbox{16}{ConsIngress} \\
        \bitbox{16}{ConsEgress} & \bitbox[lrt]{16}{}\\
        \wordbox[lrb]{1}{HopFieldMAC}
    \end{bytefield}
    \caption{Layout of a HopField.}
    \label{fig:hdr:hop}
\end{figure}

The HopField, used for hops without a reservation, is slightly changed compared to SCION and is shown in \cref{fig:hdr:hop}. This HopField is also used as the first hop field for reserved hops at segment boundaries (see \cref{sec:headerspec:boundaries}).
It contains the following fields:
\begin{description}
    \item[F (new)]
        Flyover bit.
        Indicates whether this is a hop field or a flyover hop field.
        Set to 0 for HopFields.
    \item[r (unchanged)]
        Unused and reserved for future use.
    \item[I (unchanged)]
        ConsIngress Router Alert.
        If the ConsIngress Router Alert is set, the ingress router (in construction direction) will process the L4 payload in the packet.
    \item[E (unchanged)]
        ConsEgress Router Alert.
        If the ConsEgress Router Alert is set, the egress router (in construction direction) will process the L4 payload in the packet.
    \item[ExpTime (unchanged)]
        Expiry time of a hop field.
        The field is 1-byte long, thus there are 256 different values available to express an expiration time.
        The expiration time expressed by the value of this field is relative, and an absolute expiration time in seconds is computed in combination with the timestamp field (from the corresponding info field).
    \item[ConsIngress, ConsEgress (unchanged)]
        The 16-bit interface IDs in construction direction.
    \item[HopFieldMAC (name changed)]
        6-byte MAC to authenticate the hop field.
        For details on how this MAC is calculated refer to the hop-field MAC computation of the SCION path type.\footnote{\url{https://docs.scion.org/en/latest/protocols/scion-header.html\#hop-field-mac-computation}}
\end{description}

\subsection{FlyoverHopField (new)}
\label{sec:headerspec:flyover}

\begin{figure}
    \centering
    \begin{bytefield}[boxformatting=\baselinealign,bitwidth=0.03\columnwidth]{32}
        \bitheader{0-31} \\
        \bitboxes*{1}{FrrrrrIE}& \bitbox{8}{ExpTime} & \bitbox{16}{ConsIngress} \\
        \bitbox{16}{ConsEgress} & \bitbox[lrt]{16}{}\\
        \wordbox[lrb]{1}{AggMAC}\\
        \bitbox{22}{ResID} & \bitbox{10}{BW}\\
        \bitbox{16}{ResStartOffset} & \bitbox{16}{ResDuration}
    \end{bytefield}
    \caption{Layout of a FlyoverHopField.}
    \label{fig:hdr:flyover}
\end{figure}

The FlyoverHopField shown in \cref{fig:hdr:flyover} is present if the reservation bit in the previous hop field is set to 1.
It contains the following fields:
\begin{description}
    \item[F] Flyover bit.
        Indicates whether this is a hop field or a flyover hop field.
        Set to 1 for FlyoverHopFields.
    \item[r, I, E, ExpTime, ConsIngress, ConsEgress]
        These values are the same as in the standard HopField. Note that ExpTime is the expiration time of the standard HopField, not the expiration time of the reservation.
    \item[AggMAC]
        The aggregate MAC~\cite{katz2008aggregate} (i.e., XOR) of the standard HopField MAC and the per-packet flyover MAC as described in \cref{sec:details}:
        \begin{equation}
         \fl{AggMAC} = \fl{HopFieldMAC} \oplus \fl{FlyoverMAC},
        \end{equation}
        where
        \begin{subequations}%
        \begin{align}%
         \fl{FlyoverMAC} &= \prf{A_K}{\dst \concat{}\fl{PktLen} \concat{}\ts}[:6], \label{eq:flyoverMac}\\
         \ts &= \fl{ResStartOffset} \concat{} \fl{MillisTimestamp}  \concat \fl{Counter}, \label{eq:hummingbird_ts}\\
         \dst &= \fl{DstISD} \concat{}\fl{DstAS}, \label{eq:dstaddr}\\
         \fl{PktLen} &= \fl{PayloadLen} + 4\cdot \fl{HdrLen}. \label{eq:pktlen}
        \end{align}
        \end{subequations}
        PayloadLen and HdrLen are the values from the SCION Common Header and PktLen is a 2-byte value. If an overflow occurs during the calculation of PktLen, the packet must be dropped. $A_K$ is computed as described in \cref{eq:keyderivation} and \cref{sec:headerspec:keyderivation}. The bit-layout for the input to the MAC computation (\cref{eq:flyoverMac}) is shown in \cref{fig:flyoverMacInput}.
    \item[ResID]
        22-bit Reservation ID, this allows for approximately 4 million concurrent reservations for a given ingress/egress pair.
    \item[BW]
        10-bit bandwidth field indicating the reserved bandwidth which allows for \num{1024} different values.
        The values could be encoded similarly to floating point numbers (but without negative numbers or fractions), where some bits encode the exponent and some the significant digits.
        For example, one can use 5 bits for the exponent and 5 bits for the significand and calculate the value as significand if \verb|exponent = 0| or otherwise as\\ \verb|(32 + significand) << (exponent - 1)|.
        This allows values from 0 to almost $2^{36}$ with an even spacing for each interval between powers of $2$.
    \item[ResStartOffset]
        The offset between the BaseTimestamp in the Path Meta header and the start of the reservation (in seconds).
        This allows values up to approximately 18 hours in second granularity.
    \item[ResDuration]
        Duration of the reservation, i.e., the difference between the timestamps of the start and expiration time of the reservation.
\end{description}

\begin{figure}[t]
    \centering
    \begin{bytefield}[boxformatting=\baselinealign,bitwidth=0.03\columnwidth]{32}
        \bitheader{0-31} \\
        \bitbox{16}{DstISD} & \bitbox[lrt]{16}{}\\
        \wordbox[lrb]{1}{DstAS} \\
        \bitbox{16}{PktLen} & \bitbox{16}{ResStartOffset}\\
        \bitbox{10}{MillisTimestamp} & \bitbox{22}{Counter}
    \end{bytefield}
    \caption{Layout of the input to the computation for the FlyoverMAC (see \cref{eq:flyoverMac}).}
    \label{fig:flyoverMacInput}
\end{figure}

\begin{algorithm}[t]
\caption{
  Authenticating and forwarding traffic at the ingress border router of the $i$th on-path AS~(AS~$i$). For a higher-level overview, see \cref{fig:br_high_level}. Fields inside the packet~(\fl{pkt}) are represented as \field{\fl{Field}}. \fl{\sv{}_i} is the AS-local secret value for the computation of Hummingbird authentication tags.
}
\label{alg:packet_proc}
\begin{algorithmic}[1]
\renewcommand{\algorithmicrequire}{\textbf{Input:}}
\renewcommand{\algorithmicreturn}{\textbf{Return:}}

\Require{\fl{pkt}, \fl{\sv{}_i}, \fl{K_i}}

\If{flyover bit \field{F} is set}
  \State (retFlyover, TS, PktLen) $\gets$ \textsc{FlyoverProcessing}(\fl{pkt}, \fl{\sv{}_i}) \Comment{\cref{alg:flyover_proc}}
\Else
  \State retFlyover $\gets$ \fwdbest{}
\EndIf

\If {retFlyover == \droppkt{}}
\State \textbf{drop packet}
\EndIf

\State retHf $\gets$ \textsc{StandardHfProcessing}(\fl{pkt}, \fl{K_i}) \Comment{\cref{alg:scion_proc}}

\If {retHf == \droppkt{}}
\State \textbf{drop packet}
\EndIf

\If {retFlyover == \fwdfly{}}
    \State retMonitor $\gets$ \textsc{BandwidthMonitoring}(\field{\fl{ResID}}, \field{\fl{BW}}, \fl{PktLen}) \Comment{\cref{alg:scion_proc}}

    \If {retMonitor == \fwdfly{}}
    \State \textbf{forward packet as high priority}
    \EndIf
\EndIf

\State \textbf{forward packet as best effort}
\end{algorithmic}
\end{algorithm}

\subsection{Segment boundaries}\label{sec:headerspec:boundaries}
On segment boundaries, the AS needs to process two hop fields (last HF in the first segment, first HF in the second segment).
Both HFs together specify the ingress and egress interfaces.
If there is a flyover field for this AS, it must be placed in the first segment as the first HF of the AS.

\subsection{Key Derivation}\label{sec:headerspec:keyderivation}

Since the derivation of $A_K$ is done completely by AS $K$, they are free to choose the key derivation function used in the key derivation as described in \cref{eq:keyderivation} in the main paper:
\begin{displaymath}
  A_K = \prf{\sv_K}{\resinfo_K},
\end{displaymath}

The concrete input to the key derivation is
shown in \cref{fig:kdf:resinfo},
where $\fl{ResStart} = \fl{BaseTimestamp} - \fl{ResStartOffset}$.

\begin{figure}[h]
    \centering
    \begin{bytefield}[boxformatting=\baselinealign,bitwidth=0.03\columnwidth]{32}
        \bitheader{0-31} \\
        \bitbox{16}{ConsIngress} & \bitbox{16}{ConsEgress} \\
        \bitbox{22}{ResID} & \bitbox{10}{BW} \\
        \wordbox{1}{ResStart} \\
        \bitbox{16}{ResDuration} & \bitbox{16}{0 Padding} \\
    \end{bytefield}
    \caption{Layout of $ResInfo_K$ used for the key derivation in \cref{eq:keyderivation}.}
    \label{fig:kdf:resinfo}
\end{figure}

\subsection{Packet Forwarding}
\label{sec:headerspec:forwarding}

\begin{algorithm}[t]
\caption{Reservation processing at a border router. $\Delta$ is the maximum packet age, and $\delta$ is the maximum acceptable clock skew.}
\label{alg:flyover_proc}
\begin{algorithmic}[1]

\Require \fl{pkt}, \fl{\sv{}_i}
\Ensure (fwd\_type, \fl{TS}, \fl{PktLen})

\Function{FlyoverProcessing}{\fl{pkt}, \fl{\sv{}_i}}

\State \fl{ResStart} $\gets$ \field{\fl{BaseTimestamp}} - \field{\fl{ResStartOffset}}
\State \fl{ResInfo_i} $\gets$
\field{\fl{ConsIngress}} $\|$ \field{\fl{ConsEgress}} ||
\field{\fl{ResID}} $\|$ \field{\fl{BW}} $\|$  \fl{ResStart} $\|$ \field{\fl{ResDuration}}
\Comment{\cref{fig:kdf:resinfo}}

\State $A_i \gets \prf{\sv_i}{\resinfo_i}$ \Comment{\cref{eq:keyderivation}}
\State \fl{TS} $\gets$ \field{\fl{ResStartOffset}} $\|$ \field{\fl{MillisTimestamp}} $\|$ \field{\fl{Counter}} \Comment{\cref{eq:hummingbird_ts}}
\State \fl{DstAddr} $\gets$ \field{\fl{DstISD}} $\|$ \field{\fl{DstAS}}  \Comment{\cref{eq:dstaddr}}
\State (overflow, \fl{PktLen}) $\gets$ checkedAdd(\field{\fl{PayloadLen}}, $4\cdot$\field{\fl{HdrLen}}) \Comment{\cref{eq:pktlen}}
\If {overflow}
\State \Return (\droppkt{}, \fl{TS}, \fl{PktLen})
\EndIf

\State \fl{FlyoverMAC_i} $\gets \mathsf{PRF}_{A_i}$(\fl{DstAddr} $\|$ \fl{PktLen} $\|$ TS)$[:6]$ \Comment{\cref{eq:flyoverMac}}

\State \field{\fl{AggMAC_i}} $\gets$ \fl{FlyoverMAC_i} $\oplus$ \field{\fl{AggMAC_i}} \Comment{Cand. HF MAC}
\State \fl{absTS} $\gets$ \field{\fl{BaseTimestamp}} $\|$ \field{\fl{MillisTimestamp}}

\If{$now() - absTS \notin [-\delta, \Delta + \delta ]$} \Comment{Freshness check}
  \State \Return (\fwdbest{}, \fl{TS}, \fl{PktLen})
\EndIf

\State \fl{ResExp} $\gets$ \fl{ResStart} + \field{\fl{ResDuration}}

\If{now() $\notin$ [\fl{ResStart}, \fl{ResExp}]} \Comment{Res. active check}
  \State \Return (\fwdbest{}, \fl{TS}, \fl{PktLen})
\EndIf

\State \Return (\fwdfly{}, \fl{TS}, \fl{PktLen})
\EndFunction
\end{algorithmic}
\end{algorithm}

When forwarding a packet with a FlyoverHopField at the current AS, each ingress border router executes the steps described in \cref{sec:details:dataplane} 
in addition to the packet processing of the standard SCION path type with the difference that the HopFieldMAC and the FlyoverMAC are XORed before comparing them to the value stored in the header. The FlyoverMAC is computed as shown in \cref{eq:flyoverMac}.

Before forwarding, the router then replaces the AggMAC in the Flyover with the HopFieldMAC that it computes during the verification of the FlyoverHopField. This allows the path to be reversed easily.

A high-level overview is shown in \cref{fig:br_high_level}, the detailed algorithms are specified in \cref{alg:packet_proc,alg:flyover_proc,alg:scion_proc}.

In the \emph{Reservation active check} in \cref{alg:flyover_proc}, the clock skew is not included, since this could cause problems at the boundaries of two reservations that share the same \resid{}. In particular, it may i) cause packets from expired reservations to be counted towards a new, unrelated reservation, or ii) if no adjacent reservations are assigned the same \resid{}, more traffic may be prioritized than the AS has capabilities for.

\begin{algorithm}[t]
\caption{Overview of the standard SCION border router processing steps. Providing a full description of SCION border router processing is beyond the scope of this document. This algorithm highlights the border router checks that are essential to Hummingbird's security properties. \fl{K_i} is the AS-local secret value for the computation of the SCION HopFieldMACs.
}
\label{alg:scion_proc}
\begin{algorithmic}[1]

\Require \fl{pkt}, \fl{K_i}
\Ensure fwd\_type

\Function{StandardHfProcessing}{\fl{pkt}, \fl{K_i}}

\If {hop field is expired}
    \State \Return \droppkt{}
\EndIf

\State Recompute \fl{HopFieldMAC} using \fl{pkt}, \fl{K_i}

\If {\fl{HopFieldMAC} in \fl{pkt} does not match}
    \State \Return \droppkt{}
\EndIf

\State Additional checks (segment combination allowed, \dots)

\State Update the \field{\fl{SegID}}

\If {flyover bit \field{\fl{F}} is set}
\State \field{\fl{CurrHF}} $\gets$ \field{\fl{CurrHF}} + 5
\Else
\State \field{\fl{CurrHF}} $\gets$ \field{\fl{CurrHF}} + 3
\EndIf

\State \Return \fwd{}
\EndFunction

\end{algorithmic}
\end{algorithm}

\begin{figure*}[t]
    \centering
    \usetikzlibrary{positioning}
\begin{tikzpicture}[
    processing/.style={rectangle, draw=black, inner sep=2pt, align=left},
    conn/.style={->, ultra thick},
    fo/.style={greenfte},
    be/.style={bluefte, dashed},
    dr/.style={redfte, dotted},
    io/.style={font={\bfseries}},
    lab/.style={font={\bfseries}},
]
    \node (IN) [io] {pkt in};
    \node (FO) [below right=0.2cm and 1cm of IN, processing] {\textbf{Flyover processing (Algo.~\ref{alg:flyover_proc}):}\\- re-compute \fl{FlyoverMAC}\\ - compute candidate HF MAC\\- check res. timeliness};
    \node (BE) [above right=0.2cm and 5cm of IN, processing] {\textbf{Standard SCION processing (Algo.~\ref{alg:scion_proc}):}\\- check HF MAC\\- update \fl{SegID}\\- update \fl{CurrHF}\\- \dots};
    \node (BM) [right=1.0cm of FO, processing] {\textbf{Bandwidth monitoring (Algo.~\ref{alg:policing})}\\\& (optional) duplicate\\suppression};

    \node (OUTBE) [io, right=3cm of BE] {fwd best effort};
    \node (OUTDROP) [io, below=0.7cm of OUTBE, xshift=-30pt] {drop packet};
    \node (OUTFO) [io] at ([yshift=-15pt] BM -| OUTBE) {fwd high priority};

    \draw [conn, fo] (IN) |- node[lab, anchor=north] {flyover pkt} (FO);
    \draw [conn, be] (IN) |- node[lab, anchor=south] {best-effort pkt} (BE);
    \draw [conn, be] (FO) |- node[lab, pos=0.25, anchor=east, align=right] {not timely} ([yshift=15pt]BE.south west);
    \draw [conn, fo] ([yshift=20pt] FO) -| node[lab, pos=0.80,anchor=east, align=right] {timely} ([xshift=-60pt]BE.south);

    \draw [conn, be] (BE) -- node[lab, anchor=north] {check succeeds} (OUTBE);
    \draw [conn, dr] ([xshift=50] BE.south) |- node[lab, pos=0.75, anchor=north] {check fails} (OUTDROP);
    \draw [conn, fo] (BE) -- node[lab, anchor=west, align=left] {check\\succeeds} (BM.north -| BE);
    \draw [conn, be] ([yshift=-5pt]BM) -| node[pos=0.25, lab, anchor=south] {overuse} (OUTBE);
    \draw [conn, dr] ([yshift=25pt] BM) -| node[lab, pos=0.25, anchor=south, align=right] {duplicate pkt} (OUTDROP);
    \draw [conn, fo] (BM.east |- OUTFO) -- node[lab, anchor=north, align=center] {no overuse\\no duplicate} (OUTFO);

    \node (A) [below=3cm of IN] {};
    \node (B) [right=1cm of A] {};
    \node (T) [anchor=west] at (B.east) {Flyover processing path};
    \draw [conn, fo] (A) -- (B);

    \node (A) [right=0.2cm of T] {};
    \node (B) [right=1cm of A] {};
    \node (T) [anchor=west] at (B.east) {Best-effort processing path};
    \draw [conn, be] (A) -- (B);

    \node (A) [right=0.2cm of T] {};
    \node (B) [right=1cm of A] {};
    \node (T) [anchor=west] at (B.east) {Drop path};
    \draw [conn, dr] (A) -- (B);

\end{tikzpicture}
    \caption{High-level representation of the packet-processing pipeline at the border router.}
    \label{fig:br_high_level}
\end{figure*}

\subsection{Path Reversal}
Path reversal works in the same way as standard SCION with the addition that for each FlyoverHopField, the flyover bit is set to $0$, all fields that are unique to FlyoverHopFields are removed (ResID, BW, ResStartOffset, ResDuration) to convert the FlyoverHopField to a regular HopField, and the $\fl{Seg_iLen}$ values are adjusted accordingly. This provides a valid path of the \name path type (albeit without reservations), but it can further be converted to the regular SCION path type by replacing the PathMetaHdr with the PathMetaHdr of the regular SCION path type (i.e., removing the timestamps and and converting the $Seg_iLen$ values).

\section{Additional Evaluation Results}
\label{app:additional}

\subsection{Gas Cost Evaluation}
\label{app:additional:gascost}

\begin{table*}[th]
	\centering
	\caption{Gas cost (rounded to two significant figures) for contract calls to asset and marketplace contracts, as well as transactions for atomically buying and redeeming a full path. A Negative value indicates that the caller \emph{earns} SUI because the storage rebate exceeds the transaction cost.
	}\label{tab:gas_cost_full}
	\sisetup{exponent-mode = fixed, fixed-exponent = 0}
	\begin{tabular}{@{}p{0.25cm}l S[table-format=1.5] S[table-format=1.5] S[table-format=1.5] S[table-format=-1.5] S[table-format=-1.6] S@{}}
		\toprule
		\multicolumn{2}{l}{\textbf{Contract call}} & \textbf{Computation (SUI)} & \textbf{Storage (SUI)} & \textbf{Storage rebate (SUI)} & \textbf{Total (SUI)} & \textbf{Total (USD)\textsuperscript{*}}           \\
		\midrule
		\multicolumn{2}{l}{\textbf{Asset functions}}                                                                                                                                                                \\
		                                           & issue                      & 0.00075                & 0.0046                        & 0.0025               & 0.0029                                  & 0.0035  \\
		                                           & split\_time                & 0.00075                & 0.0052                        & 0.0031               & 0.0029                                  & 0.0035  \\
		                                           & split\_bandwidth           & 0.00075                & 0.0052                        & 0.0031               & 0.0029                                  & 0.0035  \\
		                                           & fuse\_time                 & 0.00075                & 0.0031                        & 0.0051               & -0.0013                                 & -0.0016 \\
		                                           & fuse\_bandwidth            & 0.00075                & 0.0031                        & 0.0051               & -0.0013                                 & -0.0016 \\
		                                           & redeem                     & 0.00075                & 0.0045                        & 0.0051               & 0.00012                                 & 0.00014 \\
		                                           & deliver\_reservation       & 0.00075                & 0.00099                       & 0.0045               & -0.0027                                 & -0.0033 \\
		\multicolumn{2}{l}{\textbf{Market functions}}                                                                                                                                                               \\
		                                           & create\_marketplace        & 0.00075                & 0.0030                        & 0.00098              & 0.0028                                  & 0.0034  \\
		                                           & register\_seller           & 0.00075                & 0.0026                        & 0.00098              & 0.0024                                  & 0.0029  \\
		                                           & create\_listing            & 0.00075                & 0.011                         & 0.0067               & 0.0050                                  & 0.0061  \\
		                                           & buy (full)                 & 0.00075                & 0.0071                        & 0.010                & -0.0023                                 & -0.0028 \\
		                                           & buy (split bw)             & 0.00075                & 0.013                         & 0.010                & 0.0039                                  & 0.0048  \\
		                                           & buy (split time)           & 0.00075                & 0.020                         & 0.010                & 0.010                                   & 0.012   \\
		                                           & buy (split both)           & 0.00075                & 0.026                         & 0.010                & 0.016                                   & 0.020   \\
		\multicolumn{2}{l}{\textbf{Atomic buy-and-redeem}}                                                                                                                                                          \\
		                                           & 1 hop                      & 0.00075                & 0.047                         & 0.016                & 0.031                                   & 0.038   \\
		                                           & 2 hops                     & 0.00075                & 0.090                         & 0.029                & 0.062                                   & 0.076   \\
		                                           & 4 hops                     & 0.00075                & 0.18                          & 0.054                & 0.12                                    & 0.15    \\
		                                           & 8 hops                     & 0.0015                 & 0.35                          & 0.10                 & 0.25                                    & 0.30    \\
		                                           & 16 hops                    & 0.0030                 & 0.69                          & 0.20                 & 0.49                                    & 0.60    \\
		\bottomrule
		\multicolumn{6}{l}{\footnotesize \sisetup{exponent-mode=input}Computation price: \num{7.5e-07} SUI/unit; storage price: \num{7.6e-06} SUI/byte;}                                                            \\ \multicolumn{6}{l}{\footnotesize \sisetup{exponent-mode=input}SUI price: \num{1.221} USD (as of 2024-04-18 14:09 UTC); }  \\
	\end{tabular}
\end{table*}

\Cref{tab:gas_cost_full} (of which the results table in \Cref{tab:gas_cost}
is a subset) shows the cost of the contract calls for our control plane.
Since the cost is deterministic, the table does not include any measurement uncertainties.
The cost for transactions in Sui is split into three components: Computation cost, storage cost, and a storage rebate.
Computation cost is based on the complexity of the computation, which is charged according to fixed buckets of computational units.
These computational units are then converted to a value in Sui according to the \emph{computation gas price} provided in the transaction.
In our results, the values are provided based on the current reference gas price on mainnet of \num{7.5e-07} SUI per unit. Storage of objects is charged based on a \emph{storage gas price}, which is currently  \num{7.6e-06} SUI per byte.
The third component is a storage rebate, which the transaction sender receives when deleting items from the storage, and amounts to 99\% of the original value paid in SUI for the storage of the item. If the storage rebate exceeds the transaction's computation and storage cost, the sender receives the difference.

As \cref{tab:gas_cost_full} shows, the cost of transactions for individual calls for interacting with the assets or the marketplace are relatively cheap, costing only fractions of a cent.
The cost of buying a full path depends on the length of the path, and it is dominated by the cost of buying multiple assets, since this incurs the storage cost of splitting an asset and re-listing the pieces that are not bought.
In our benchmark, each asset on a path requires a worst-case split, i.e., the buyer buys a time interval from the middle of the time interval represented by the listing and only a fraction of the bandwidth. This causes two splits in the time dimension and one in the bandwidth dimension.

Since buying an asset creates new objects and delivering a reservation deletes them, most of the storage fee incurred by the buyer is later refunded \emph{to the AS} as a storage rebate.
To lower the transaction cost for the buyer, the market could be pre-loaded with funds provided by the AS, which then the smart contract could use to refund part of the cost of buying assets.
The AS later gets these funds back through the storage rebate when deleting the assets.

\subsection{Border Router Processing}
\label{app:additional:processing}

\Cref{fig:dp_eval_router_latency} reports the execution times of all the steps required to process a regular SCION packet, and the additional overhead incurred when processing a \name packet (darker gray background). A SCION packet requires \SI{123}{ns} to be processed, while \name adds an overhead of \SI{185}{ns} for a total of \SI{308}{ns}.

\begin{table}[t!]
	\centering
	\caption{Fine-grained packet validation and forwarding timings at the border router.
		The timings are independent of the number of on-path ASes and the size of the payload.}
	\begin{tabular*}{\linewidth}{l@{\extracolsep{\fill}} S[table-format=2.0]} 
     \toprule
     \textbf{Task} & \textbf{Time [ns]}\\ 
     \midrule
     Check packet size & 14 \\
     Parse packet headers & 30 \\
     Check whether hop field is expired & 8 \\
     Recompute SCION hop field MAC & 46 \\
     Update segment identifier (SegID) & 4 \\
     Update current hop field pointer & 13 \\
     Check if hop field is of type SCION or Flyover & 8 \\
     \rowcolor{gray!15}
     Compute absolute start of reservation (ResStart) & 8 \\
     \rowcolor{gray!15}
     Compute authentication key ($A_i$) & 43 \\
     \rowcolor{gray!15}
     AES-extend authentication key ($A_i$) & 24 \\
     \rowcolor{gray!15}
     Validate high-precision time stamp & 6 \\
     \rowcolor{gray!15}
     Recompute flyover MAC & 44 \\
     \rowcolor{gray!15}
     Compute aggregate MAC & 4 \\
     \rowcolor{gray!15}
     Verify xor-ed MAC same as in header & 9 \\
     \rowcolor{gray!15}
     Check whether the reservation is still active & 8 \\
     \rowcolor{gray!15}
     Check for overuse & 39 \\
     \textbf{Total} & \textbf{123 \colorbox{gray!15}{[+185]}} \\
     \bottomrule
\end{tabular*}
	\label{fig:dp_eval_router_latency}
\end{table}

\subsection{Traffic Generation Performance}
\label{app:additional:trafficgeneration}

To complete our data-plane benchmarks, we study the performance of a \name \emph{traffic generator}.
In most cases, we expect the sources to be end-hosts, where the forwarding performance is limited by the hosts' stacks and network uplinks.
However, performant traffic generation is essential in scenarios where a single entity purchases bandwidth for hosts within its network---for example, a corporate LAN---and provides reservations to hosts through a dedicated \name gateway.

\Cref{fig:dp_eval_sender_tp_c} shows the traffic-generation throughput of a multi-core \name{} gateway implemented in DPDK.
Increasing the number of cores enhances the throughput:
For payloads of \SI{500}{\byte}, a mere \num{32} cores deliver \SI{160}{Gbps} line rate for both \name{} and SCION, even across long paths with eight on-path ASes.
\Cref{fig:dp_eval_sender_latency} details the durations of individual per-packet computation steps for a four-hop path, where the total overheads are \SI{494}{ns} and \SI{293}{ns} for \name{} and SCION, respectively.

Traffic generation at the source is notably slower than traffic forwarding at the border router.
This is as expected: The source has to compute the \name packet authentication tags for all ASes on the path, while on-path border routers only need to compute the authentication tag for their own AS. Nevertheless, a \num{32}-core source gateway is sufficient to serve the vast majority of deployments, as residential and corporate link speeds seldom exceed \SI{1}{Gbps}, and even ISP links are usually below~\SI{100}{Gbps}.

\begin{figure}[t]
	\input{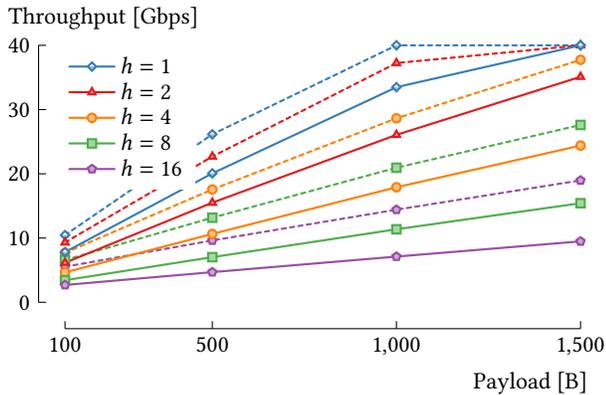}
	\caption{Single-core packet generation performance at the source for different number of AS-level hops ($h$) and payloads.
		Solid lines correspond to Hummingbird reservations, dashed lines to standard SCION best-effort traffic.}
	\label{fig:dp_eval_sender_tp_pl}
\end{figure}

\Cref{fig:dp_eval_sender_tp_pl} demonstrates the single-core packet generation performance at the source.
For both SCION and Hummingbird traffic, the throughput achieved correlates with the number of ASes on the path and scales proportionally to the packets' payload size.
For instance, with a \SI{1}{\kilo\byte} payload size and flyovers on four on-path ASes, \name{} achieves a throughput of \SI{17.90}{Gbps}, whereas SCION's best-effort traffic reaches \SI{28.64}{Gbps}.
However, with smaller \SI{100}{\byte} payloads, the throughput decreases to \SI{4.65}{Gbps} and \SI{7.70}{Gbps}, respectively.

\begin{figure}[t]
\definecolor{eth1}{HTML}{1F407A}
\definecolor{eth2}{HTML}{3C5A0F}
\definecolor{eth3}{HTML}{0069B4}
\definecolor{eth4}{HTML}{72791C}
\definecolor{eth5}{HTML}{91056A}
\definecolor{eth6}{HTML}{6F6F6E}
\definecolor{eth7}{HTML}{A8322D}
\definecolor{eth8}{HTML}{007A92}
\definecolor{eth9}{HTML}{956013}
\definecolor{eth10}{HTML}{82BE1E}

\pgfplotsset{compat=1.17}
\pgfplotscreateplotcyclelist{eval list source}{
    bluefte, dashed, every mark/.append style={solid,fill=bluefte!50},mark=halfsquare*\\
	redfte, dashed, every mark/.append style={solid,fill=redfte!50},mark=triangle*\\
	yellowfte, dashed, every mark/.append style={solid,fill=yellowfte!50},mark=*\\
	greenfte, dashed, every mark/.append style={solid,fill=greenfte!50},mark=square*\\
        purplefte, dashed, every mark/.append style={solid,fill=purplefte!50},mark=pentagon*\\
   bluefte, every mark/.append style={solid,fill=bluefte!50},mark=halfsquare*\\
	redfte, every mark/.append style={solid,fill=redfte!50},mark=triangle*\\
	yellowfte, every mark/.append style={solid,fill=yellowfte!50},mark=*\\
	greenfte, every mark/.append style={solid,fill=greenfte!50},mark=square*\\
    purplefte, every mark/.append style={solid,fill=purplefte!50},mark=pentagon*\\
}

\begin{tikzpicture}

\begin{axis}
[
font=\small,
xlabel={Number of CPU cores},
ylabel={Throughput [\si{Gbps}]},
legend entries={,,,,,$h = 1$, $h = 2$, $h = 4$, $h = 8$, $h = 16$},
xmin=1, xmax=32,
ymin=0, ymax=160,
xtick={1, 2, 4, 8, 16, 32},
ytick={0, 40, 80, 120, 160},
height=5cm, width=\axisdefaultwidth,
legend style={anchor=south east, at={(1.01,0)}, cells={anchor=west}, row sep=-2pt},
legend columns=2,
cycle list name=eval list source,
]

\addplot
table {%
1 26.00
2 51.94
4 103.62
8 159.76
16 160
32 160
};

\addplot
table {%
1 22.61
2 45.14
4 89.04
8 159.78
16 160
32 160
};

\addplot
table {%
1 17.52
2 35.02
4 69.47
8 127.69
16 160
32 160
};

\addplot
table {%
1 13.12
2 26.27
4 52.12
8 95.52
16 160
32 160
};

\addplot
table {%
1 9.61
2 19.23
4 38.11
8 69.53
16 121.57
32 160
};

\addplot
table {%
1 19.97
2 40.05
4 79.41
8 146.64
16 160
32 160
};

\addplot
table {%
1 14.98
2 30.77
4 61.19
8 112.48
16 160
32 160
};

\addplot
table {%
1 10.38
2 21.41
4 42.36
8 77.59
16 135.96
32 160
};

\addplot
table {%
1 7.01
2 14.02
4 27.77
8 50.78
16 88.60
32 160
};

\addplot
table {%
1 4.68
2 9.36
4 18.52
8 33.79
16 59.13
32 118.06
};

\end{axis}
\end{tikzpicture}
	\caption{Packet generation performance at the source for packets carrying a payload of \SI{500}{\byte}, and for different number of AS-level hops ($h$) and cores.
		Solid lines correspond to Hummingbird reservations, dashed lines to standard SCION best-effort traffic.}
	\label{fig:dp_eval_sender_tp_c}
\end{figure}

\begin{table}[t!]
	\centering
	\caption{Fine-grained packet generation timings at the source, for four AS-level hops; the additional operations required for Hummingbird are highlighted. All timings are independent of the payload size except in the case of ``Add packet payload'', which we evaluate
		for \SI{500}{\byte} and \SI{1500}{\byte} payloads.
	}

\centering
\begin{tabular}{lrr} 
     \toprule 
    \textbf{Task} & \multicolumn{2}{S}{\textbf{Time [ns]}}\\ 
     \midrule
     Add Ethernet, IP, Scion header fields & 
     107 \\ 
     \rowcolor{gray!15}
     Compute flyover MACs (4 on-path ASes) & 
     201 & \\
     Add hop fields for all on-path ASes &  
     171 \\
     Add \SI{500}{\byte} (\SI{1500}{\byte}) payload  & 15 & (40) \\
     \textbf{Total for \SI{500}{\byte} (\SI{1500}{\byte}) payload} & \bfseries 494 & \bfseries (519) \\
     \bottomrule
\end{tabular}
	\label{fig:dp_eval_sender_latency}
\end{table}

\section{Bidirectional Reservation Support}\label{sec:bidirectional}

Bidirectional reservations can generally be implemented by having the source communicate pre-computed hop authentication tags to the destination with every packet.
If the reverse traffic is expected to be larger, multiple tags per hop can be provided in each packet.
However, this approach has two main problems:
(i) sending additional authentication tags for each hop on the path in almost every packet introduces a large overhead, and (ii) this approach increases the complexity of monitoring, since it requires aggregating the monitoring information for forwards and backwards traffic, which are monitored at two different border routers.

We believe that bidirectional reservations are better obtained with a separate---and possibly out-of-band---exchange protocol:
\begin{itemize}
	\item The source obtains reservations to the destination normally and obtains separate reservations for the reverse path.
	\item The source communicates the authentication keys to the destination for the reverse path (if the destination is enabled to use the reservations).
	\item Source and destination use the reservations as normal.
\end{itemize}

Note that, even though on the data plane both source and destination use unidirectional “forward” reservations, these are both billed to the source and therefore act as a backward reservation.
This is an important property enabled by the control-plane independence of \name:
The source can obtain a reservation in any direction and for any hop, which is not the case in previous bandwidth-reservation protocols such as Colibri~\cite{giuliari2021colibri} or Helia~\cite{wyss2022protecting}.

With our control plane, the source could even send the bandwidth assets for the backward reservation to the destination (assuming it knows their on-chain identity), which would allow the destination to directly obtain the reservation information and authentication keys by automatically redeeming the assets.
In this case, it may be useful to use additional tags in the asset that contain the source address to ensure that the (trusted) destination uses the reservations for communicating with the source.

\end{document}